\documentclass[useAMS,usenatbib]{mn2e}
\usepackage{amssymb}
\usepackage[dvips]{graphicx}
\usepackage{color}

\def\pmi{$\pm$}

\title[FR~Cnc Revisited: Photometry, Polarimetry and Spectroscopy]{FR~Cnc Revisited: Photometry, Polarimetry and Spectroscopy\thanks{Based on the
 observations made: with the 2.2-m telescope of the
German-Spanish Astronomical Centre, Calar Alto (Almer\'{\i}a,
Spain), operated by Max-Planck-Institute for Astronomy,
Heidelberg, in cooperation with the Spanish National Commission
for Astronomy; with the Nordic Optical Telescope (NOT), operated
on the island of La Palma jointly by Denmark, Finland, Iceland,
Norway and Sweden, in the Spanish Observatorio del Roque de Los
Muchachos of the Instituto de Astrof\'isica de Canarias; with the
Isaac Newton Telescope (INT) operated on the island of La Palma by
the Isaac Newton Group in the Spanish Observatorio del Roque de
Los Muchachos of the Instituto de Astrof\'isica de Canarias; with
the Italian Telescopio Nazionale Galileo (TNG) operated on the
island of La Palma by the Centro Galileo Galilei of the INAF
(Istituto Nazionale di Astrofisica) at the Spanish Observatorio
del Roque de Los Muchachos of the Instituto de Astrof\'isica de
Canarias; with ASAS-3 survey; with robotic 0.35-m telescope at the
Sonoita Research Observatory (Arizona, USA); with 29-cm telescope,
operated by Terskol Branch of the Astronomy Institute, Russia;
with 104-cm Sampurnanand Telescope of ARIES, Nainital, India; with
2.0-m Himalayan Chandra Telescope, operated at the Indian
Astronomical Observatory (Mt. Saraswati, Hanle, India).} }
\author[A. Golovin et~al.]{A. Golovin$^{1}$\thanks{E-mail:
golovin.alex@gmail.com}, M.C. G\'alvez-Ortiz$^{2, 3}$, M.
Hern\'an-Obispo$^4$, M. Andreev$^{1, 5, 6}$, \newauthor J.R.
Barnes$^{3}$, D. Montes$^4$, E. Pavlenko$^7$, J.C. Pandey$^8$, R.
Mart\'inez-Arn\'aiz$^4$,
\newauthor B.J. Medhi$^8$, P.S. Parihar$^9$, A. Henden$^{10}$, A. Sergeev$^{5, 6}$,  S.V. Zaitsev$^1$, N. Karpov$^{5, 6}$ \\
$^{1}$Main Astronomical Observatory of  National Academy of Sciences of Ukraine,  Zabolotnogo str., 27, Kiev, 03680, Ukraine \\
$^{2}$Centro de Astrobiolog\'ia (CSIC-INTA). Crta, Ajalvil km 4.
E-28850
Torrej\'on de Ardoz, Madrid, Spain \\
$^{3}$Centre for Astrophysics Research, University of Hertfordshire, College Lane, Hatfield, Hertfordshire AL10 9AB, UK \\
$^{4}$Astrophysics department, Physic Faculty, Universidad Complutense de Madrid, E-28040 Madrid, Spain \\
$^{5}$Terskol Branch of the Astronomy Institute of RAS, Kabardino-Balkaria Republic, 361605, Russia \\
$^{6}$International Center for Astronomical, Medical and
Ecological Research of National Academy of \\ Sciences of Ukraine (ICAMER of NASU), Zabolotnogo str., 27, Kiev, 03680, Ukraine \\
$^{7}$Crimean Astrophysical Observatory (CrAO), Nauchny, 98409, Ukraine \\
$^{8}$Aryabhatta Research Institute of Observational Sciences
(ARIES),
Manora Peak, Nainital, 263129, India \\
$^{9}$Indian Institute of Astrophysics, Block II, Koramangala,
Bangalore, 560 034, India \\
$^{10}$AAVSO, Clinton B. Ford Astronomical Data and Research
Center, 49 Bay State Rd. Cambridge, MA 02138, USA}

\begin{document}
\date{Accepted 1988 December 15. Received 1988 December 14; in original form 1988 October 11}

\pagerange{\pageref{firstpage}--\pageref{lastpage}} \pubyear{2009}

\maketitle

\label{firstpage}

\begin{abstract}
This is a part of a multiwavelength study aimed at use of
complementary photometric, polarimetric and spectroscopic data to
achieve an understanding of
 the activity process in late-type stars.
Here we present the study of FR~Cnc, a young, active and spotted
star.

We performed analysis of $ASAS-3$ (The All Sky Automated Survey)
data for the years 2002--2008 and amended the value of the
rotational period to be 0.826518 d. The amplitude of photometric
variations decreased abruptly in the year 2005, while the mean
brightness remained the same, which was interpreted as a quick
redistribution of spots. $BVR_{c}$ and $I_c$ broad band
photometric calibration was performed for 166 stars in FR~Cnc
vicinity.

The photometry at Terskol Observatory shows two brightening
episodes, one of which occurred at the same phase as the flare of
2006 November 23. Polarimetric $BVR$ observations indicate the
probable presence of a supplementary source of polarization. We
monitored FR~Cnc spectroscopically during the years 2004--2008. We
concluded that the RV changes cannot be explained by the binary
nature of FR~Cnc. We determined the spectral type of FR~Cnc as
K7V. Calculated galactic space-velocity components ($U,~V,~W$)
indicate that FR~Cnc belongs to the young disc population and
might also belong to the IC~2391 moving group. Based on Li~{\sc
i}~$\lambda$6707.8 measurement, we estimated the age of FR~Cnc to
be between 10--120 Myr. Doppler Tomography was applied to create a
starspot image of FR~Cnc. We optimized the goodness of fit to the
deconvolved profiles for axial inclination, equivalent width and
$v$~sin~$i$, finding $v$~sin~$i$~=~$46.2$~km~s$^{-1}$ and
$i~=~55^{\circ}$. We also generated a synthetic $V$-band
lightcurve based on Doppler imaging that makes simultaneous use of
spectroscopic and photometric data. This synthetic lightcurve
displays the same morphology and amplitude as the observed one.

The starspot distribution of FR~Cnc is also of interest since it
is one of the latest spectral types to have been imaged. No polar
spot was detected on FR~Cnc.
\end{abstract}


\section{Introduction}

It is well-known that late-type stars show magnetic activity
similar to the activity of our Sun, but the physics of `stellar'
activity is not yet well understood. In addition, the activity
level manifested by late-type stars is much higher than that
observed for the Sun.

FR~Cnc (= BD+16~1753 = MCC~527 = 1ES~0829+15.9 =
1RXS~J083230.9+154940 = HIP~41889) was first mentioned as a
probable active star when it was identified as the optical
counterpart of a soft X-ray source 1ES~0829+15.9 in the
\emph{Einstein Slew Survey}. It has $V~=~10.43$~mag, spectral type
K8V, the X-ray flux of $\approx~10^{-11}~\rmn{erg~s^{-1}~cm^{-2}}$
(\citeauthor{intr2} \citeyear{intr2}; \citeauthor{intr9}
\citeyear{intr9}). Lately, this object was rediscovered as an
X-ray source 1RXS~J083230.9+154940 in the \emph{ROSAT} All-Sky
Survey (RASS) with lower X-ray flux at the level of
$2~\times~10^{-12}~\rmn{erg~s^{-1}~cm^{-2}}$ (\citeauthor{intr11}
\citeyear{intr11}). The X-ray luminosity of
$(2~-~12)~\times~10^{29}~\rmn{erg~s^{-1}}$ and the ratio of X-ray
to bolometric luminosity $\frac{F_{x}}{F_{bol}}$ of $\geq
10^{-3.3}$ (\citeauthor{intr7} \citeyear{intr7}) indicates that
this object has an active corona (\citeauthor{intr9}
\citeyear{intr9}).

In the \emph{Hipparcos} catalogue this star was mentioned as an
unsolved variable star with the identifier HIP~41889 and
$0.17$~mag amplitude of variability (\citeauthor{intr1}
 \citeyear{intr1}). It was classified as BY~Dra type star (i.e. its
variability caused by rotational modulation of starspots) and
given the name FR~Cnc by \citet{intr3}. For analysis of
\emph{Hipparcos} observations see \citet{intr7}. FR~Cnc
($\rmn{RA}(2000) = 08^{\rmn{h}} 32^{\rmn{m}} 30\fs5287$ and
$\rmn{Dec.}~(2000) = +15\degr 49\arcmin 26\farcs193$) has
$30.24~\pm~2.03$ marcsec parallax (\citeauthor{intr1}
\citeyear{intr1}) that implies a distance of $33~\pm~2$~pc and an
absolute magnitude of $7.8$. The kinematics of FR~Cnc suggests
that it is a very young (35--55 Myr) main-sequence star and a
possible member of the IC 2391 supercluster, as it was shown by
\citet{intr7}. \citet{intr10} concluded that FR~Cnc is not a
binary system, based on two measurements of the RV.

The presence of Ca~{\sc ii}~H \& K and H$\alpha$ emission lines in
the spectra indicates high chromospheric activity in FR~Cnc
(\citeauthor{intr5} \citeyear{intr5}; \citeauthor{intr6}
\citeyear{intr6}). In 'quiescent' state this object manifests
optical variability with the dominant period $0.8267~\pm~0.0004$~d
due to the presence of starspots and axial rotation
(\citeauthor{intr7} \citeyear{intr7}). In addition, photometry
obtained in 2005 February -- April with Kilodegree Extremely
Little Telescope indicates FR~Cnc optical variability with
$0.827$~d period (\citeauthor{pepper}
 \citeyear{pepper})
 when monitoring the Praesepe open cluster for
transiting exoplanets.

\begin{figure}
\includegraphics [width=8cm]{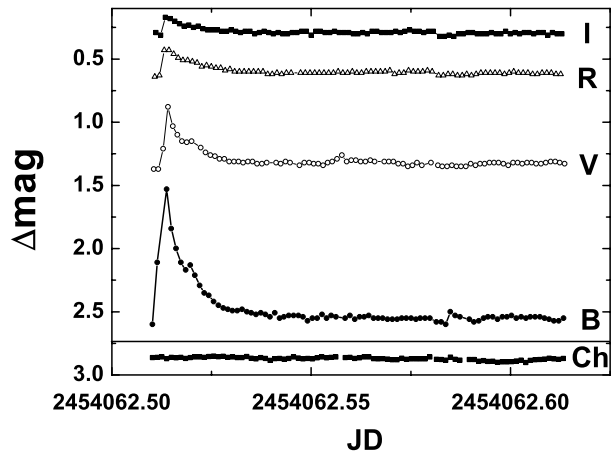}
 \caption{Lightcurve of FR~Cnc on 2006 November 23 from \citet{golovin}: "The flare of FR~Cnc: shifted
differential lightcurves in $B-, V-, R-$ and $I-$bands as well as
the difference \emph{check star${}-{}$comparison star} (`Ch' on
the plot)"} \label{IBVS}
\end{figure}

The first ever-detected optical flare of this object was observed
 during CCD photometry of FR~Cnc on 2006 November 23 at Crimean Astrophysical Observatory (Ukraine) with 38-cm Cassegrain telescope
and described by \citet{golovin}. The flare was observed in
$BVRI$-bands (see Fig. \ref{IBVS}). The amplitude  reached even 1
mag in the $B$-band and was decreasing towards the $I$-band. The
flare energy output in the $B$-band was about $1.73\times10^{31}
\rmn{erg}~\AA^{-1}$ ~and flare to quiescent flux ratio was 38.63
per cent.

\section[]{Photometric Observations}

Most of the information on the photospheric activity (e.g.
starspots) of BY Dra-type stars comes from photometric
observations. The mean brightness level is strongly dependent on
the percentage of spotted area of the surface, while changes in
spot distribution over the surface could result in changes of the
amplitude of variability. FR~Cnc has a short (for such class of
objects) rotational period of 0.8267 d. As was shown in
\citet{dorren}, this short rotational period leads us to expect
large flare activity of the star.

The detection of a flare on FR~Cnc on 2006 November 23 motivated
us to continue photometric monitoring of this object as well as to
study its archival \emph{ASAS-3} observations (The All Sky
Automated Survey; see \citet{asas} for description of equipment
and data pipeline).

\subsection[]{ASAS Photometrical Observations} \label{21}

\begin{figure}
\includegraphics [width=8cm]{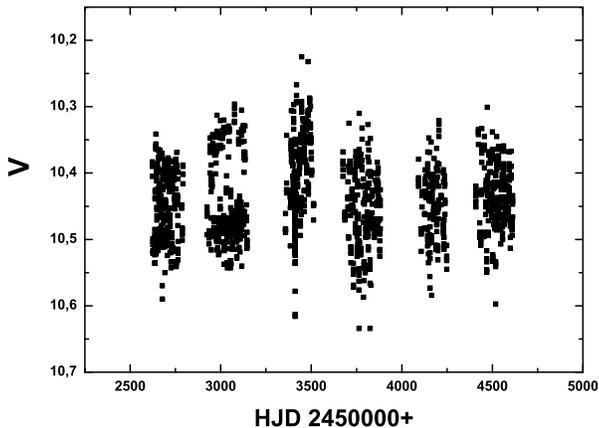}
 \caption{ASAS long term light curve}
 \label{asasLT}
\end{figure}

\begin{figure*}
\includegraphics [width=13cm]{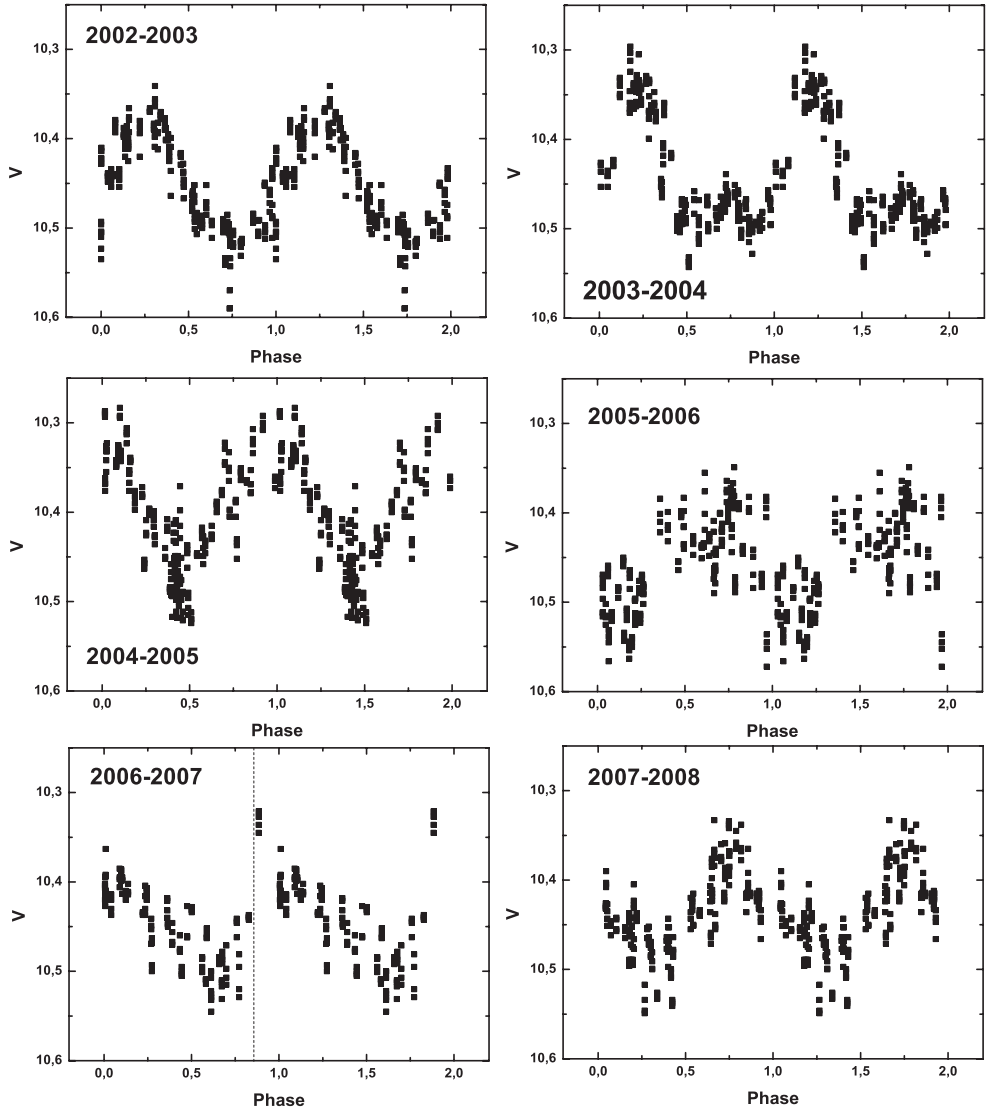}
 \caption{\emph{ASAS-3} phase diagrams for six observing seasons. Please note that scales are the same for all plots. Vertical dashed line indicates the phase where the flare on 2006 November 23 was detected (see text for the explanation).}
 \label{asas_phase}
\end{figure*}

FR~Cnc was observed in $V$-band with \emph{ASAS-3} survey  during
2002 December -- 2008 May (6 observational seasons, see
Fig.~\ref{asasLT}). All the data were split into separate data
sets according to the 'seasonal gaps' in observations and folded
with the $0.8267$ d period (from \citeauthor{intr7}
\citeyear{intr7}) and represented in Fig. \ref{asas_phase}
(plotted twice for clarity). Table \ref{table:asaslog} represents
the log of observations.

The initial epoch was common to calculate phases for all 6 phase
diagrams and was chosen arbitrary as HJD~(UTC)~=~2452635.72669
(first point in dataset). No evidence of flares in the
\emph{ASAS-3} data was found. The vertical dashed line on
Fig.~\ref{asas_phase} indicates the phase when the flare on 2006
November 23 occurred.

\emph{ASAS-3} data are not covering flare on 2006 November 23.
Nearest \emph{ASAS-3} observations were done on JD~2453881 and
JD~24544091, what is 181 days before and 29 days after the flare.
\emph{ASAS-3} observations in 2006-2007 at a phase $\sim$0.88 (143
days after the flare) show that FR~Cnc was brighter than during
the rest of the time. This probably could be related to
brightening episode, which was detected at the same phase during
Terskol observations (see sect. \ref{Tersk}).

\begin{figure}
\includegraphics [width=8cm]{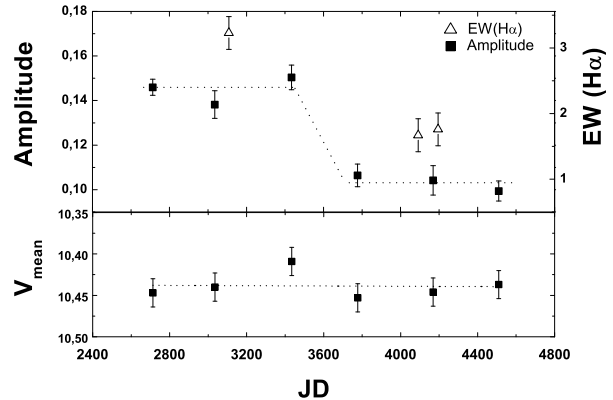}
 \caption{Mean amplitude of variability, $EW$ of H$\alpha$ and mean brightness of FR~Cnc during the years 2002--2008. (See sect. \ref{Ha} regarding H$\alpha$.)}
 \label{asas_amplitude}
\end{figure}

Profiles of variability as well as the amplitude of variability
are different from season to season, while the mean brightness
remains constant ($V_{\rmn{mean}} = 10.439$ mag within the error
limits of $\sigma = 0.017$ mag). To illustrate this, we plotted
the amplitude and the mean brightness as a function of mean epoch
of observations (Fig.~\ref{asas_amplitude}). An abrupt decrease of
amplitude in the year 2005 is clearly seen. One of the possible
interpretations is continuous spottedness of the star and
redistribution of spots/spot groups from season to season: i.e.
spots didn't disappear, but distributed more uniformly over FR~Cnc
surface. To support this idea, notice the constancy of the mean
brightness level. If spots disappear then increase of brightness
and decrease of amplitude could be expected.

\subsubsection{Periodogram Analysis}
The data set was searched for periodic variation of brightness (in
order to estimate with better accuracy the known rotational
period) using the {\sc period04} package, developed by Patrick
Lenz (Institute of Astronomy, University of Vienna, Austria; see
\citeauthor{period04} \citeyear{period04}). Discrete Fourier
transform (DFT) algorithm was applied for statistical analysis.
Julian dates are heliocentrically corrected. The average zero
point of $10.439$ mag was subtracted to prevent the appearance of
additional features on the periodogram centered at frequency 0.0.

Making a periodicity analysis of \emph{ASAS-3} photometry, we
found a dominant frequency of $f~=~1.209895$~$\rmn{c~d^{-1}}$,
while, as defined by equation \ref{eq1}, $\sigma(f)~=~0.000022$
(therefore, period P~=~$0.826518~\pm~0.000015$ d; see
Fig.~\ref{periodogram}). The detected periodicity could be
interpreted as the rotational period and it is in good agreement
with the period founded by \citet{intr7}, but with improved
accuracy (due to the longer time span of \emph{ASAS-3}
observations). No other periodicity of FR~Cnc brightness
modulations was found on the basis of \emph{ASAS-3} observations.

The DFT routine was applied separately to each season of observations
as well. The obtained periodograms do not reveal any other
significant periodicity, but only the same peak as for the periodogram
for the whole time-string. The dominant frequency remains constant
within the error limits of $\sigma(f) = 0.0003$ for all of 6
periodograms.

\subsubsection{Estimation of Accuracy and Reliability of the Detected Period}

Empirical results from observational analysis
(\citeauthor{breger93} \citeyear{breger93}) and numerical
simulations (\citeauthor{Kuschnig} \citeyear{Kuschnig}) have shown
that the ratio in amplitude in the periodogram between signal and
noise should not be lower than 4.0 to give good confidence in the
detected peak. We calculated the S/N-ratio from the periodogram
for the determined dominant frequency, so S/N = 6.88.

\begin{center}
\begin{figure*}
\includegraphics[width=8cm]{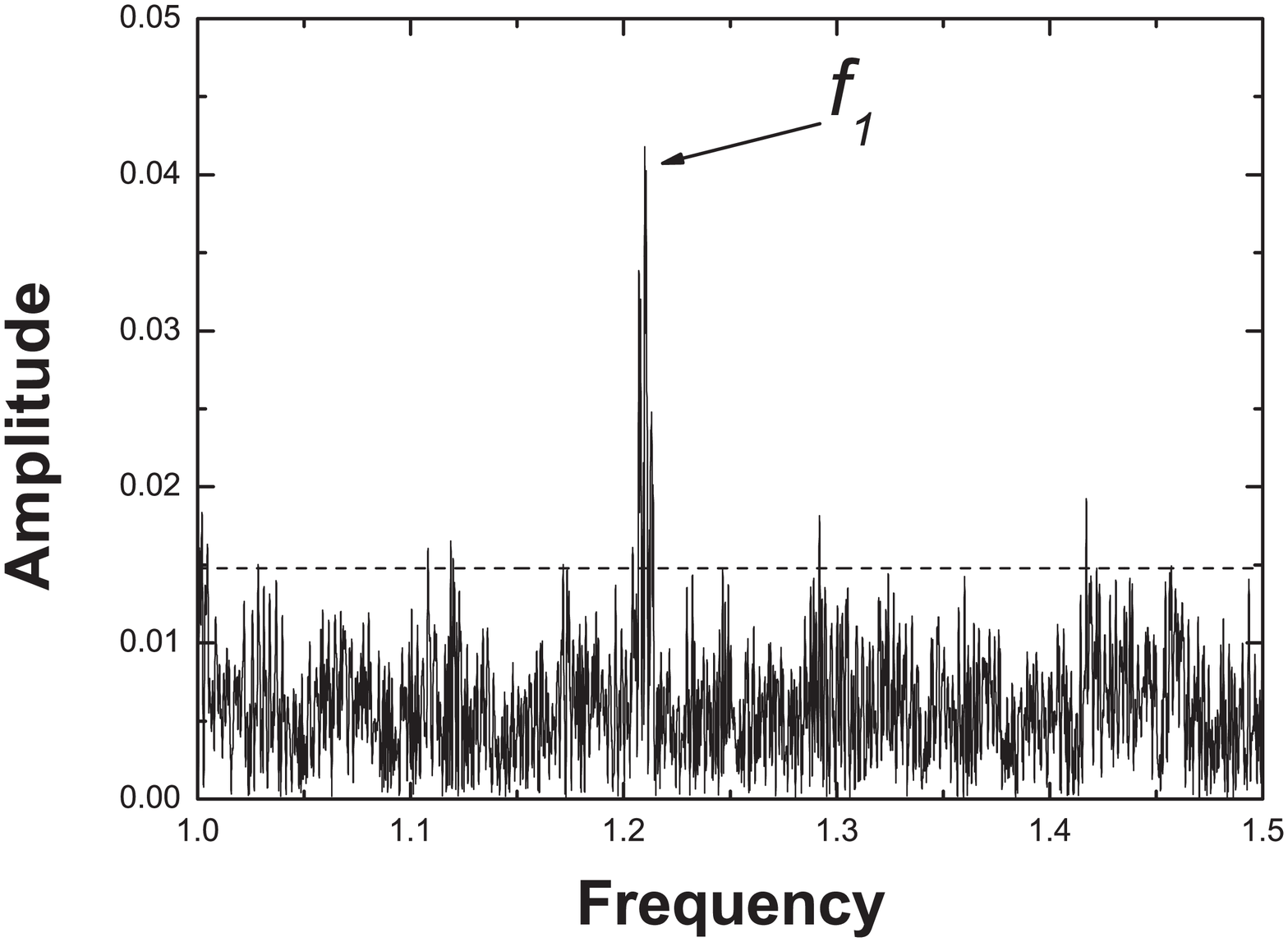}
\includegraphics[width=8cm]{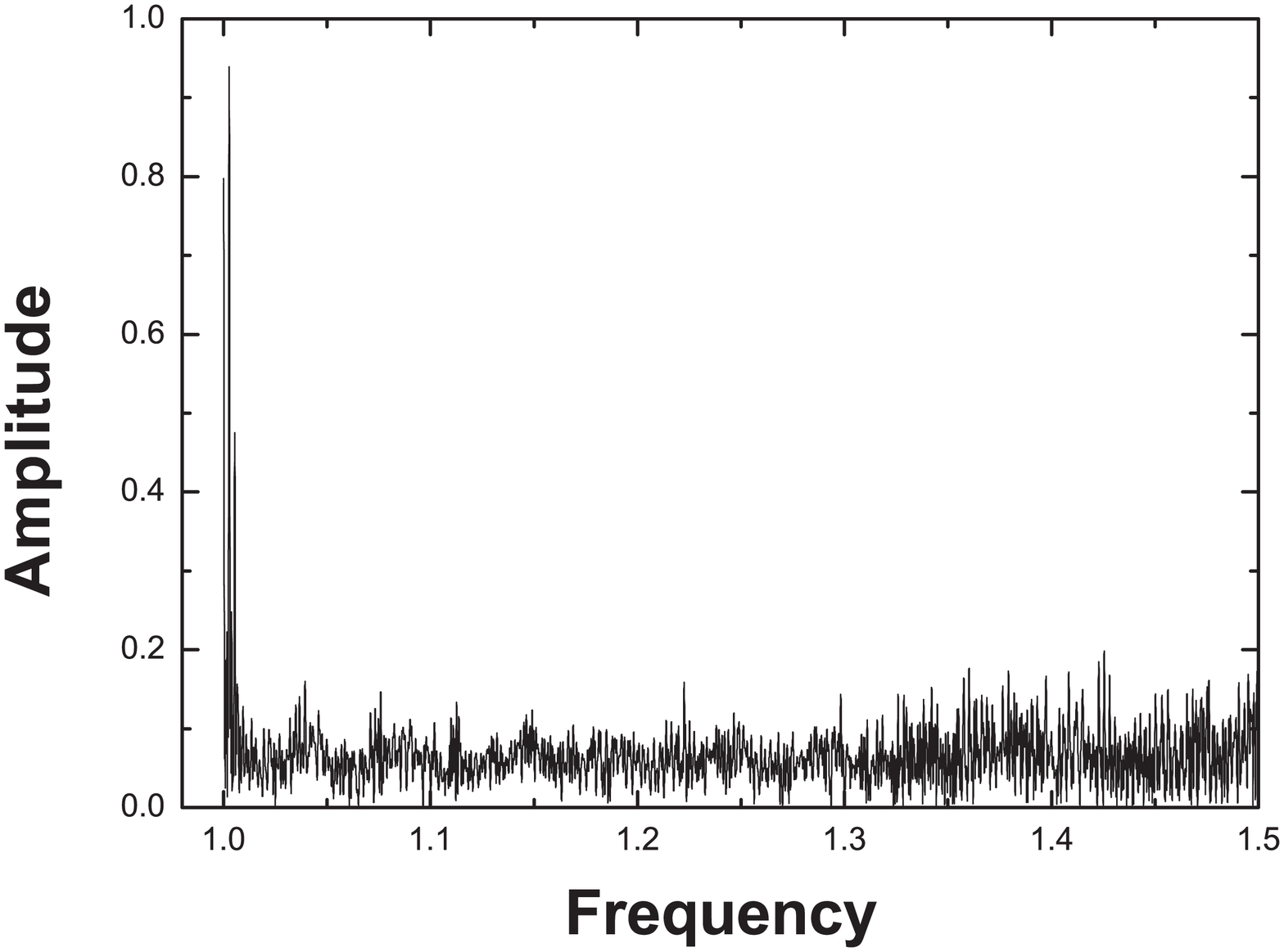}
\caption{\label{periodogram} Periodogram of \emph{ASAS-3} data in
frequency range of 1.0--1.5 and Spectral Window}
\end{figure*}
\end{center}

The Spectral Window Function is important to confirm that the
obtained frequencies are real or an artefact of the window
function. This was done by assigning 1 to brightness values with
the same observation times and checking the resulting diagram.
There was no evidence of significant power at the location of the
peak (Fig.~\ref{periodogram}). The dominant frequency on the
spectral window is $F = 1.0027$ with amplitude $A = 0.9391$, which
is due to daily gaps in observations.

Parameter uncertainties were calculated from an error matrix,
which is a by-product of non-linear least-squares fitting
procedure. Other types of uncertainties are those which could be
calculated from analytically derived formulae assuming an ideal
case. Based on some assumptions one can derive a formula for the
uncertainties in frequency and signal amplitude at this frequency.
See \citet{sigma1} and \citet{sigma2}, \citet{sigma3} for the
derivation based on a monoperiodic fit. The determined equation
can be applied for each frequency separately:
\begin{equation} \label{eq1}
\sigma(f) = \sqrt{\frac{6}{N}} \frac{1}{\pi T}\frac{\sigma (m)}{a}
\end{equation}
\begin{equation} \label{eq2}
\sigma(a) = \sqrt{\frac{2}{N}} \sigma(m)
\end{equation}
where $N$ is the number of time points, $T$ is the time length of
the data set, $\sigma(m)$ denotes the residuals from the fit and
\emph{a} refers to the signal amplitude at the particular
frequency. Both 'analytical' and 'least-squares error matrix'
calculations give similar results: $\sigma(f) = 0.000022$ and
$\sigma(a) = 0.0037$. So, on the periodogram
(Fig.~\ref{periodogram}) we plotted the $3\sigma(a)$-level (dashed
line) to show that the detected peak exceeds it significantly.

Another method for 1$\sigma$-level of amplitude calculation was
proposed for use by R.A. Fisher (see \citeauthor{fisher}
\citeyear{fisher}; \citeauthor{fisher2} \citeyear{fisher2}) and
often called as Fisher Randomization Test. The idea is to take the
original light curve and preserving the time column, shuffle the
corresponding intensities around. That destroys any coherent
signal in the light curve while keeping the time sampling intact.
We are left with a shuffled light curve of pure white noise. The
next step is to compute a DFT of this light curve which will look
really noisy. The standard deviation of the average amplitude of
such a DFT is close to the 1$\sigma$ limit. We iterated this 25
times, therefore we use the average of 25 standard deviation
values to determine 1$\sigma$-level. This pipeline gave us the
value of $\sigma(a) = 0.0028$, which is slightly less than
$\sigma(a)$ from 'analytical' and 'least-squares error matrix'
calculations, hence we plotted the bigger value on our periodogram
to be confident.

\begin{table}
\scriptsize
\begin{center}
\caption{\label{table:asaslog} Log of \emph{ASAS-3}-observations
of FR~Cnc}
\begin{tabular}{cccc}
\hline Year & $T_{\rmn{start}}$  & $T_{\rmn{end}}$ & $N_{\rmn{points}}$ \\
 & 2450000+ & 2450000+ & \\
\hline

2002--2003 &  2635  &  2791  &  59 \\
2003--2004 &  2924  &  3146  &  62 \\
2004--2005 &  3357  &  3512  &  63 \\
2005--2006 &  3674  &  3881  &  61 \\
2006--2007 &  4091  &  4247  &  34 \\
2007--2008 &  4409  &  4611  &  56 \\
\hline
\end{tabular}
\end{center}
\end{table}

\subsection[]{Photometric Sequence}

We carried out the $BVR_{c}I_{c}$ photometric calibration for 166
stars in the vicinity of FR~Cnc with $V$-magnitudes in the range
$9.85$--$18.06$ mag that could serve as comparison stars.
Calibration was done at the Sonoita Research Observatory (Arizona,
USA) using a robotic 0.35-m telescope, equipped with an SBIG
STL-1001XE CCD camera. A table with this photometrical sequence is
available electronically only via the AAVSO
ftp-server\footnote{ftp://ftp.aavso.org/public/calib/frcnc.dat}.
For user convenience, we have used the {\sc aladin} Sky Atlas
(\citeauthor{aladin} \citeyear{aladin}) to align our calibration
data on a DSS2/STScI POSSII
image\footnote{http://www.mao.kiev.ua/ardb/ref/agolovin.html}.

\subsection[]{Terskol Observations} \label{Tersk}

Optical $B$-band photometry was carried out from 2007 March  to
2008 February  at Terskol Branch of the Astronomy Institute
(Russia) with 29-cm telescope and Apogee-47 Alta CCD camera. All
observations were made in the $B$-band as the flare amplitude is
expected to increase with decreasing wavelength. The duration of
each observing run varies from 2 to 7 h. See Table
\ref{terskollog} for log of observations. The calibration process
of the obtained frames, comparison and check stars remains the
same as described by \citet{golovin}.

\begin{figure*}
\includegraphics [width=10cm]{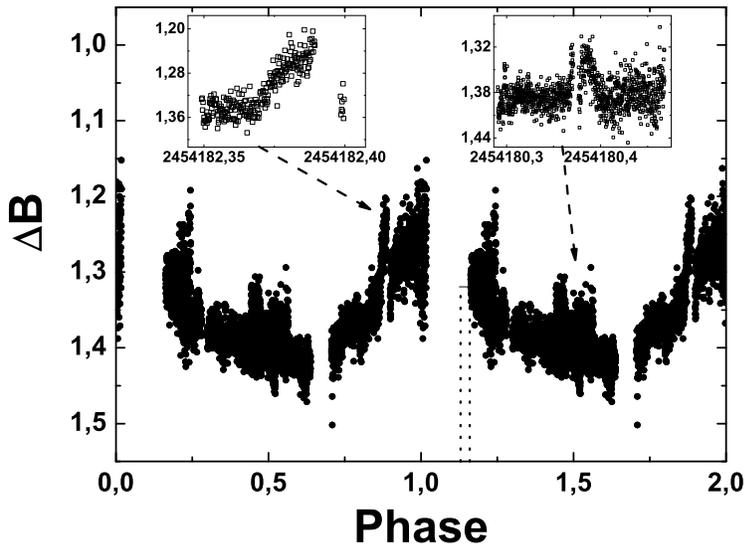}
\caption{$B$-band Photometry at Terskol Observatory in 2007 March.
Plotted twice for clarity. Dotted lines denote phases when
polarimetric observations were done. Note an enlarged plots of
brightening episodes on JD = 2454180 and JD = 2454182.}
\label{photometry2007}
\end{figure*}

The mean amplitude of FR~Cnc brightness variations in the $B$-band
during the observations in the year 2007 was $0.12$ mag and $0.13$
mag in the year 2008. There are no peculiar features in the light
curve from the year 2008, while in the photometry obtained in the
year 2007 two brightening episodes were detected: namely, on
2454180.3 and 2454182.3 (see Fig.~\ref{photometry2007}). Amplitude
in $B$-band was $0.06$ and $0.12$ mag respectively. It has to be
noted that the second episode (HJD = 2454182.3) occurred at the
same phase as the flare of 2006 November 23 (phase = $0.88$).
Probably both the events (flare on 2006 November 23 and
brightening episode at phase 0.88) are originated from the same
long-living active regions on the surface of FR~Cnc.

\begin{table}
\scriptsize
\begin{center}
\caption{\label{terskollog} Log of Terskol photometric
observations of FR~Cnc in the years 2007--2008}
\begin{tabular}{ccc}
\hline Year & Beginning of the Run &  End of the Run \\
 & (HJD) & (HJD) \\
\hline
2007 March & 2454171.2453 & 2454171.4361   \\
 & 2454174.3501 & 2454174.4019  \\
 & 2454180.1863 & 2454180.4691   \\
 & 2454182.2395 & 2454182.4972   \\
 & 2454188.4424 & 2454188.4721   \\
\hline 2008 February & 2454498.4150 & 2454498.5233  \\
 & 2454500.2675 & 2454500.5345  \\
 & 2454501.1575 & 2454501.5403  \\
 & 2454502.1681 & 2454502.5870  \\
 & 2454503.2157 & 2454503.4481  \\
 & 2454504.1755 & 2454504.3025  \\
  \hline
\end{tabular}
\end{center}
\end{table}

\section[]{Polarimetric Observations} \label{polarimetry}

The $BVR$ broad-band polarimetric observations of FR~Cnc were
obtained on 2007 October 19 and 20 using TK 1024 pixel$^2$ CCD
camera mounted on the Cassegrain focus of the 104-cm Sampurnanand
Telescope of ARIES, Nainital (India). The optical imaging
polarimetry was carried out in $B, V$ and $R$ ($\lambda_{B_{\rm
eff}}$ = 0.440 $\rm{\mu}$m, $\lambda_{V_{\rm eff}}$ = 0.550 $\mu$m
and $\lambda_{R_{\rm eff}}$ = 0.660 $\mu$m,) photometric bands.

\begin{table*}
\scriptsize
\begin{center}
\caption{\label{table:polar1} Observed polarized standard stars.}
\begin{tabular}{llllll}
\hline Star name & Filter  & $P$\pmi$\epsilon_{P}$$(per cent)$ &
$\theta$\pmi $\epsilon_{\theta}$$(^\circ)$ &
$P$\pmi$\epsilon_{P}$$(per cent)$ &
$\theta$\pmi$\epsilon_{\theta}$$(^\circ)$ \\
\hline
 & & \multicolumn{2}{c}{Published data}&\multicolumn{2}{c}{This paper}\\
\hline
HD 25433          &B & 5.23\pmi0.09 & 134.3\pmi 0.05  & 5.17\pmi 0.21&  135.6 \pmi 1.0 \\
                  &V & 5.12\pmi0.06 & 134.2\pmi 0.03  & 5.13\pmi 0.09&  133.5 \pmi 0.8 \\
                  &R & 4.73\pmi0.05 & 133.6\pmi 0.03  & 4.76\pmi 0.13&  132.9 \pmi 0.5 \\
HD 19820           &B & 4.70\pmi0.04 & 115.70\pmi 0.22 & 4.66\pmi 0.07& 115.49 \pmi 0.19 \\
                  &V & 4.79\pmi0.03 & 114.93\pmi 0.17 & 4.76\pmi 0.10& 114.15 \pmi 0.20 \\
                  &R & 4.53\pmi0.03 & 114.46\pmi 0.17 & 4.56\pmi 0.17& 114.18 \pmi 0.21 \\
\hline
\end{tabular}
\end{center}
\end{table*}

Standard stars for null polarization and for the zero-point of the
polarization position angle were taken from \citet{schmidt}. The
results for standards are given in Table \ref{table:polar1}. From
the results, it can be concluded that obtained values of
polarization and position angles are in good agreement with
\citet{schmidt} within the error limit.

Both the program and the standard stars were observed during the
same night. HD 25433 and HD 19820 (= CC Cas) were used as a
standard polarized stars, while HD 21447 and G 191-B2B (= HIP
23692) served as standard unpolarized stars.

The results are listed in Tables \ref{table:polar1},
\ref{table:polar2} and \ref{table:polar3}, where $P$ is the
fraction of the total light in the linearly polarized condition
and $\theta$ is the position angle of polarization plane to the
equatorial plane. It  is denoted by the normalized Stokes'
parameter ${\it q}\ (=Q/I)$, when the half wave plate's fast axis
is aligned to the reference axis ($\alpha = 0^\circ$). Similarly,
the normalized Stokes' parameter ${\it  u}\ (=U/I)$, when  the
half wave plate is at 22.5$^\circ$. For further details on used
equipment and the method of observations, refer to \citet{medhi}.

By dotted lines on the lightcurve we indicate the phases
($\varphi_1 = 0.16, \varphi_2 = 0.13$ for 2007 October 19 and 20
respectively), when our polarimetric observations were conducted
(see Fig.~\ref{photometry2007}). FR~Cnc was in maximal brightness
during that time.

Polarization in FR~Cnc could be magnetic in origin. The degree
of polarization depends  nonlinearly on the size of  magnetic
regions (see \citeauthor{polar1} \citeyear{polar1};
\citeauthor{polar2}
 \citeyear{polar2}).
 These authors have also calculated
a grid of expected degrees of polarization in $UBVRI$ band for
stars with temperature from 4000 to 7000~K and log~$g$ from 2.0 to
4.5. We have used their results to compare our observed values of
polarization for FR~Cnc. Fig.~\ref{fig:model} represents the
degree of polarization for FR~Cnc in $BVR$ bands. The maximum
possible degree of polarization for the total spot area of 24 per
cent is derived from the calculations of \citet{polar2} for the
star corresponding to the spectral type of {K5--7V} and
characteristic magnetic field of 2.7~kG. These values are
over-plotted and represented by a solid line in
Fig.~\ref{fig:model}. The observed polarization in the $B$ band is
in good agreement with the theoretical values expected for Zeeman
polarization model. However, the observed polarization in $V$ and
$R$ bands slightly exceeds the theoretical values.

\begin{figure}
\includegraphics [width=8cm]{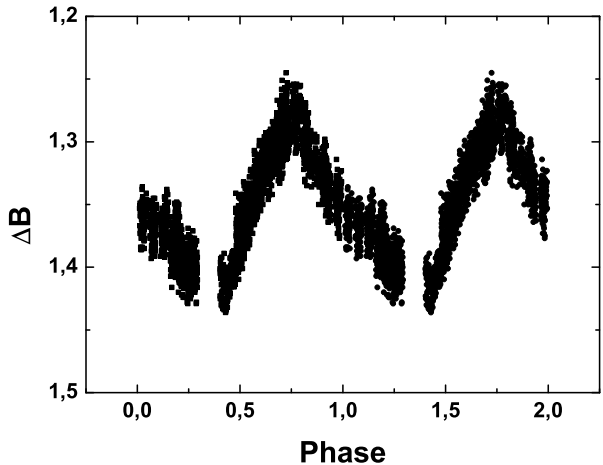}
\caption{$B$-band Photometry at Terskol Observatory in 2008
February.} \label{photometry2008}
\end{figure}

Model values for K2V-K7V and K2IV-K2V spectral types appear to be
even lower and certainly do not match the polarization in any of
observed bands. This is also observed in some other young spotted
stars (MS Ser, LQ~Hya, VY~Ari; see \citeauthor{alekseev}
 \citeyear{alekseev}) and probably due to the presence of a
supplementary source of linear polarization.

The predicted values of polarization due to Thompson and Rayleigh
scattering from inhomogeneous regions are not enough to explain
the observed polarization excess (Thompson and Rayleigh scattering
for the assumed spectral type supposed to be of order of $10^{-7}$
and $10^{-4}$ per cent respectively; \citeauthor{polar2}
\citeyear{polar2}). The mechanism which can produce additional
linear polarization is probably scattering in circumstellar
material (e.g. see \citeauthor{Pandey09} \citeyear{Pandey09}); on
the other hand, the mentioned models are unacceptable if FR~Cnc is
a close binary star (\citeauthor{alekseev} \citeyear{alekseev};
\citeauthor{elias} \citeyear{elias} and \citeauthor{polar2}
\citeyear{polar2}).

\begin{table}
\scriptsize
\begin{center}
\caption{\label{table:polar2} Observed unpolarized standard stars}
\begin{tabular}{llll}
\hline Star name & Filter  & $q$ & $u$ \\
\hline
HD 21447          &B & 0.019 & 0.011 \\
                  &V & 0.037 & -0.031 \\
                  &R & -0.035 & -0.039 \\
G191B2B        &B & 0.072& -0.059 \\
                  &V & -0.022 & -0.041 \\
                  &R & -0.036 & 0.027 \\
\hline
\end{tabular}
\end{center}
\end{table}

\begin{table*}
\scriptsize \caption{\label{table:polar3} Observed $BVR$
polarization values for FR~Cnc.}
\begin{tabular}{lllll}
\hline Date of Observation & Filter  & Time (UT) & $P$\pmi$\epsilon_{P}$$(per cent)$ & $\theta$\pmi $\epsilon_{\theta}$$(^\circ)$\\
\hline
October 19, 2007 &B & 22:22:12.0 & 0.22 \pmi 0.05 & 57 \pmi 7 \\
                  &V & 22:01:45.1 & 0.20 \pmi 0.02 & 55 \pmi 2 \\
                  &R & 22:10:52.6 & 0.16 \pmi 0.05 & 61 \pmi 7 \\
October 20, 2007 &B & 21:25:36.6 & 0.26 \pmi 0.04 & 55 \pmi 5 \\
                  &V & 21:30:00.5 & 0.24 \pmi 0.07 & 54 \pmi 6 \\
                  &R & 21:35:48.6 & 0.17 \pmi 0.05 & 58 \pmi 7 \\
\hline
\end{tabular}
\end{table*}

\begin{figure}
\includegraphics[width=8cm]{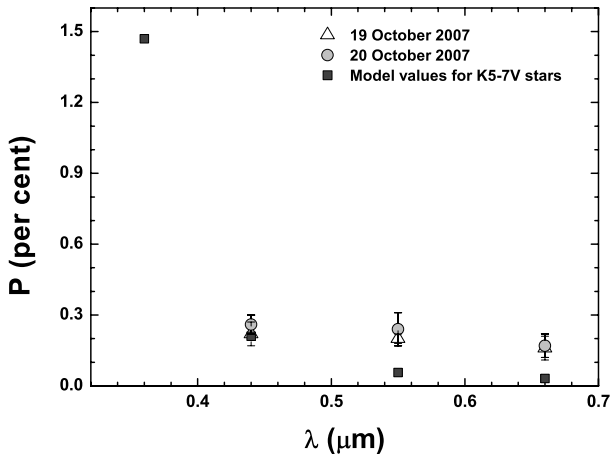}
\caption{The degree of polarization of FR~Cnc as a function of
wavelength. Model values of polarization for K5-7V spectral types
stars (\citeauthor{polar2}, \citeyear{polar2}) are plotted for
comparison.} \label{fig:model}
\end{figure}

\section[]{Spectroscopic Observations}
\begin{table*}
\caption[]{Description of spectroscopic observations
\label{tab:obs}}
\begin{flushleft}
\begin{center}
\scriptsize
\begin{tabular}{clclllcccc}
\noalign{\smallskip} \hline \noalign{\smallskip} Number & Date &
Telescope & Instrument & Detector & Spect. range & Orders &
Dispersion & FWHM$^{1}$ & S/N \\
  & (dd/mm/yyyy) &           &      &  & (\AA) & & (\AA) & (\AA/pixel) & H$\alpha$ \\
\noalign{\smallskip} \hline \noalign{\smallskip}
1 & 29/03--07/04/2004 & 2.2-m$^{a}$ & FOCES$^{d}$ & 2048x2048 24$\mu$ SITE$\#1$d & 3450--10700 & 112 & 0.04--0.13  & 0.08--0.035 & 40\\
2 & 11--13/04/2004    & 2.0-m$^{b}$ & HFOSC$^{e}$ & 2000x4000 SiTe ST-002, grism Gr14 & 3270--6160 & 1 & 3.57 & 7.23--1.47  & 200\\
  &                &             &             & 2000x4000 SiTe ST-002, grims Gr8  & 5800--8350 & 1  & 3.23 & 3.66--3.99  & 200  \\
3 & 16--21/12/2006  & 2.2-m$^{a}$ & FOCES$^{d}$ & 2048x2048 24$\mu$ SITE$\#1$d & 3600--10700 & 106 & 0.08--0.1   & 0.08--0.04  & 40\\
4 & 24--26/02/2007 & 2.2-m$^{a}$ & FOCES$^{d}$ & 2048x2048 24$\mu$ SITE$\#1$d & 3600--10700 & 106 & 0.04--0.13  & 0.07--0.41  & 40\\
 & 07--08/05/2007 & & & & & & & & \\
5 & 21/03/2008   & NOT$^{c}$   & FIES$^{f}$  & 2000x2000 EEV42-40  & 3620--7360  & 80  & 0.02--0.04  & 0.05--0.11  & 80\\
\noalign{\smallskip} \hline \noalign{\smallskip}
\end{tabular}
\end{center}
$^{1}$ Full Width at Half Maximum of the arc comparison lines; \\
$^{a}$ 2.2-m telescope at the German-Spanish Astronomical
Observatory (CAHA, Almer\'{\i}a, Spain);\\ $^{b}$ 2.0-m Himalayan
Chandra Telescope at the Indian Astronomical Observatory (Mt.
Saraswati, Hanle, India); \\ $^{c}$ Nordic Optical Telescope (NOT)
at the Observatorio del Roque de los Muchachos (La Palma, Spain);
\\ $^{d}$ The Fibre Optics Cassegrain Echelle Spectrograph
(FOCES);\\ $^{e}$ Himalaya Faint Object Spectrograph and Camera
(HFOSC);\\
$^{f}$ The high-resolution Fibre-fed Echelle Spectrograph (FIES).\\
\end{flushleft}
\end{table*}

A total of 58 high and low resolution spectra of FR~Cnc have been
obtained and analysed in this work. The spectroscopic data were
obtained during five observing runs. Details of each observing run
are given in Table \ref{tab:obs}: date, telescope, spectrograph,
CCD chip, spectral range covered, number of orders included in
each echelle spectrum, range of reciprocal dispersion, spectral
resolution (determined as the full width at half maximum, FWHM, of
the arc comparison lines) and mean S/N in the H$\alpha$ line
region.

The spectra were extracted using the standard reduction procedures
in the {\sc iraf}\footnote{{\sc iraf} is distributed by the
National Optical Observatory, which is operated by the Association
of Universities for Research in Astronomy, Inc., under contract
with the National Science Foundation.}
  echelle package (bias subtraction, flat-field division and
 optimal extraction of the spectra). We obtained the wavelength
 calibration by taking spectra of a Th-Ar lamp. Finally, we normalized
 the spectra by a polynomial fit to the observed continuum.

\section[]{Stellar Parameters}
Stellar parameters of FR~Cnc are given in Table~\ref{tab:par} and
Table~\ref{tab:par2}. The photometric data ($B-V,~V$), $P_{\rm
phot}$, projected rotational velocity ($v\sin{i}$), and galactic
space-velocity components ($U,~V,~W$) have been determined in this
paper. The astrometric data (parallax, $\pi$; proper motions,
$\mu$$_{\alpha}$$\cos{\delta}$ and $\mu$$_{\delta}$) are from
\emph{Hipparcos} and \emph{Tycho-2} catalogues (\citeauthor{ESA}
 \citeyear{ESA}).

\begin{table*}
\caption[]{Stellar parameters of FR~Cnc \label{tab:par}}
\begin{center}
\scriptsize
\begin{tabular}{cccccccccccccc}
\noalign{\smallskip} \hline \noalign{\smallskip}
  {T$_{\rm sp}$} & $V_{\rm mean}$ &
 {$B-V_{\rm mean}$} & {$P_{\rm phot}$} &
{\it v}~sin~{\it i} & $i$ \\
    &   &        & \scriptsize (d) &
\scriptsize (km $~$s$^{-1}$) & $^{\circ}$ \\
\noalign{\smallskip} \hline \noalign{\smallskip}
 K7V & 10.44 & 1.10 & 0.826518$\pm$0.000015 & 46.2$\pm$0.8 & 55 \\
\noalign{\smallskip} \hline \noalign{\smallskip}
\end{tabular}
\caption[]{Astrometric and kinematic parameters of FR~Cnc
\label{tab:par2}} \scriptsize
\begin{tabular}{cccccccccccccc}
 \hline
\noalign{\smallskip}
 $\pi$ & $\mu$$_{\alpha}$ cos $\delta$ &
$\mu$$_{\delta}$ &
 $U\pm \sigma_{\it U}$ & $V \pm \sigma_{\it V}$ & $W \pm \sigma_{\it W}$ & $V_{\rm Total}$\\
(marcsec) & (marcsec yr$^{-1}$) & (marcsec yr$^{-1}$) &(km $~$s$^{-1}$) & (km$~$s$^{-1}$) & (km$~$s$^{-1}$)   & (km $~$s$^{-1}$) \\
\noalign{\smallskip} \hline \noalign{\smallskip}
 30.24$\pm$2.03$^\ddag$ &  -98.1$\pm$1.6$^\ddag$ & -91.0$\pm$1.5$^\ddag$  & -19.02$\pm$0.62 & -18.99$\pm$1.08 & -8.03$\pm$1.58  & 28.05 \\
\noalign{\smallskip} \hline \noalign{\smallskip}
\end{tabular}
 \\ $^{\ddag}$ From $Hipparcos$ and $Tycho-2$ catalogues (\citeauthor{ESA} \citeyear{ESA}).\\
\end{center}
\end{table*}

\subsection[]{Spectral classification} \label{51spectral}
FR~Cnc is classified as a K8V-star by \citet{intr9}, while a
multicolour photometric study allowed  \citet{intr7} to classify
it as a K5V-star. They also have obtained the Spectral Energy
Distribution (SED) of the star matching a $T_{\rm eff}$ of
4250$\pm$250 K and a log~$g$ of 4.50$\pm$0.5, that agrees with a
K5V classification.

We have compared our high resolution echelle spectra, in several
spectral orders free of lines sensitive to chromospheric activity,
with spectra of inactive reference stars of different spectral
types and luminosity classes, observed during the same observing
run. This analysis makes use of a modified version of the program
{\sc starmod} ({\sc jstarmod}) developed at Penn State University
(\citeauthor{Barden} \citeyear{Barden}; \citeauthor{lopez10}
\citeyear{lopez10}). This program constructs a synthesized stellar
spectrum from artificially rotationally broadened, radial-velocity
shifted, and weighted spectra of appropriate reference stars. For
FR~Cnc, we have obtained the best fit with a K7V reference star,
which is in closer agreement with the K8V classification rather
with the K5V from \citet{intr7}.

\subsection[]{Radial Velocity} \label{52}
We have determined the heliocentric radial velocities (RV
hereafter) by making use of
 cross-correlation technique (see e.g. \citeauthor{Galvez07} \citeyear{Galvez07}).
 The spectra of the
 target were cross-correlated order by order, using the routine
{\sc fxcor} in {\sc iraf}, against spectra of RV standards with
similar spectral type taken from \citet{Beavers}. We derived the
RV for each order from the position of the peak of the
cross-correlation function (CCF) and calculated the uncertainties
 based on the fitted peak height and the antisymmetric noise as
described by \citet{TonryDavis}.

\begin{figure}
\includegraphics [width=8cm]{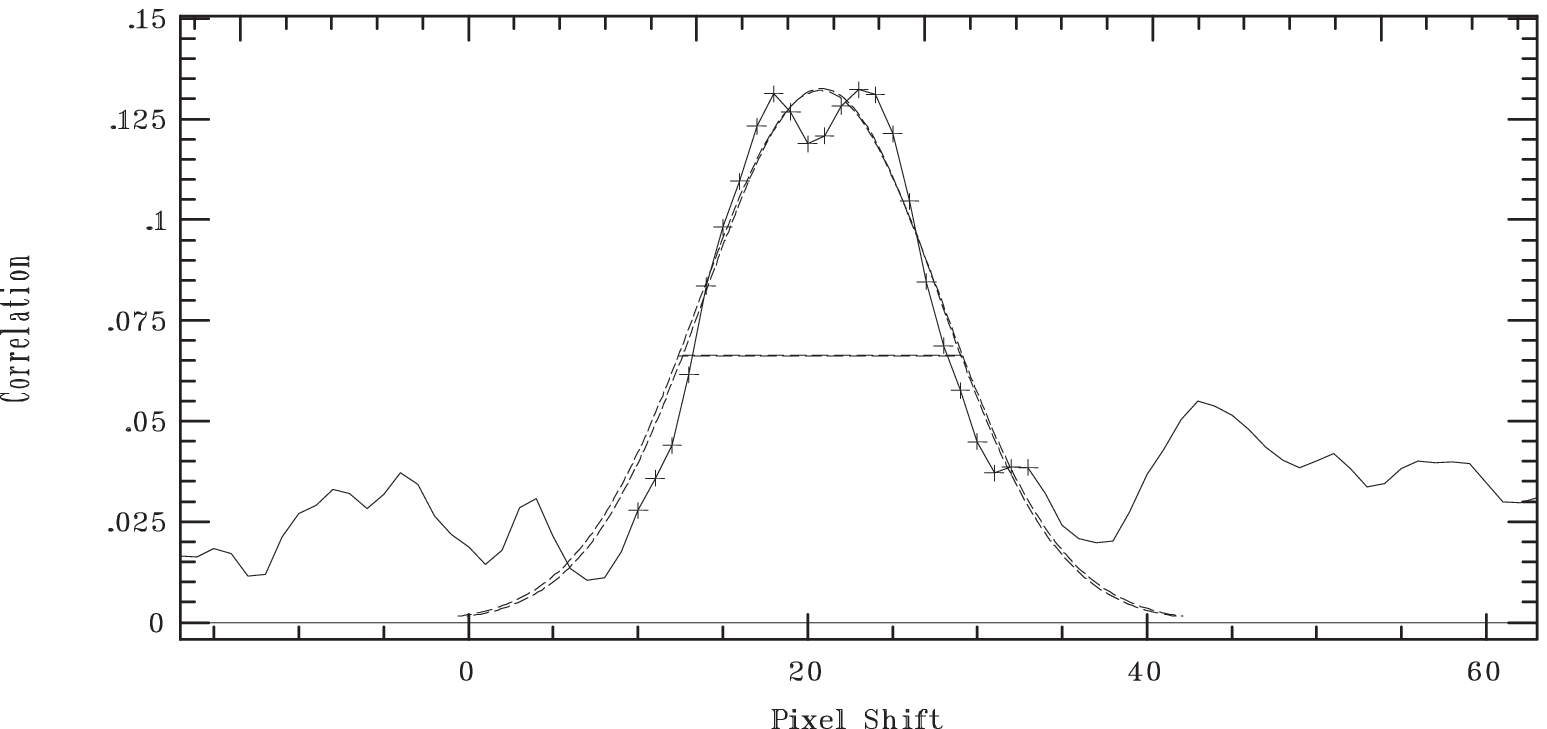}
\includegraphics [width=8cm]{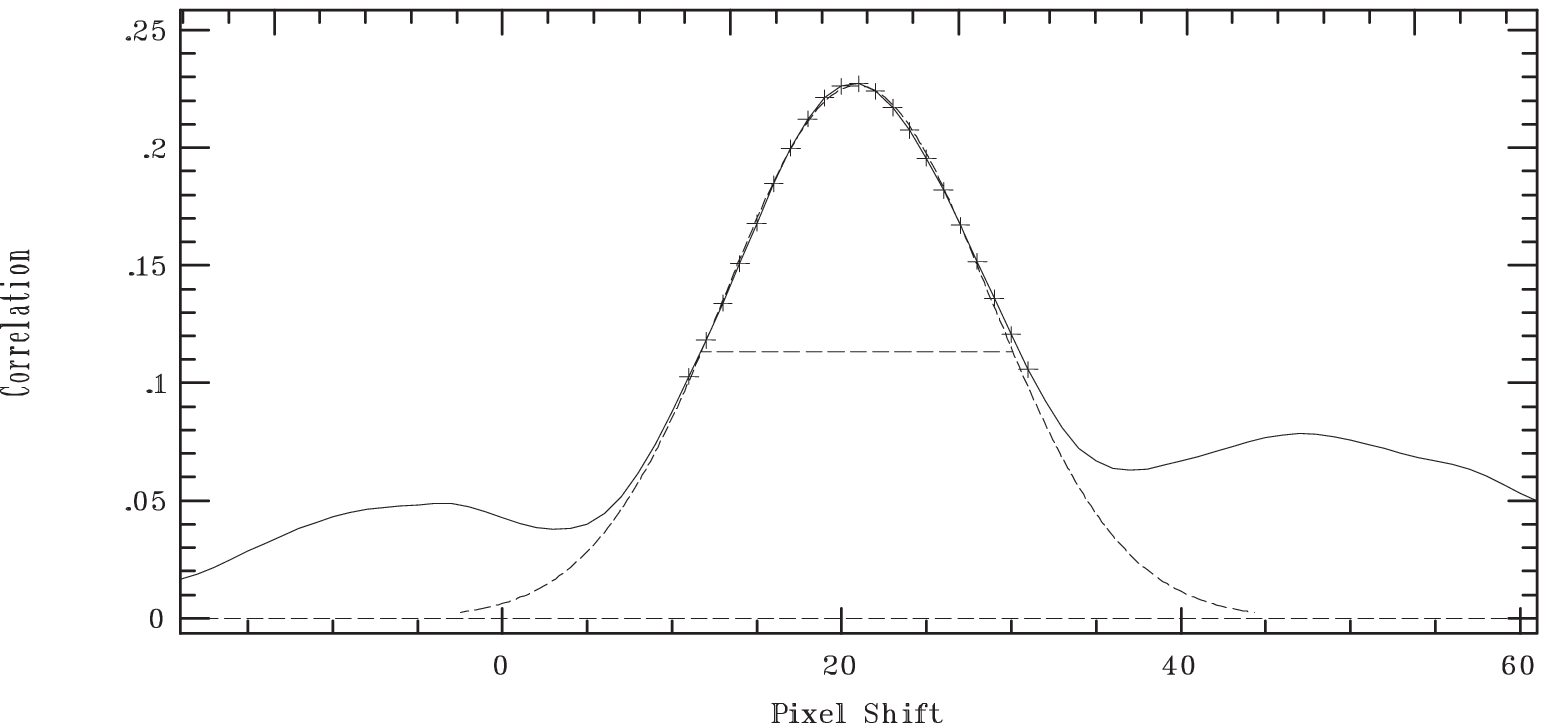}
 \caption{(Top) An example of CCF of FR~Cnc in FOCES04 observing run.
 Irregular profiles can be seen in the peak. These
 irregularities can produce significant errors in RV
 determination. (Bottom) The same CCF obtained when
 we broadened the standard star to FR~Cnc rotational velocity.
 Irregular profiles become smoother and could be
 fitted with a Gaussian.}
 \label{fig:perfil}
\end{figure}

As Fig.~\ref{fig:perfil} (top) shows, the irregular profiles of
the CCF (double peaks and asymmetries) can produce significant
errors in RV measures. These features show regular variations: a
double peak moving on time-scale of the
 rotational period can be seen. Photospheric activity features on the stellar
 surface that disturb the profile of the photospheric lines
 could induce variations in the peak of the CCF, but a stellar companion could also produce the double peak effect.

 We checked if these RV variations could be
 due to a binary nature of FR~Cnc. We found no evidence for the existence of a companion by measuring the RV of the peaks and by trying to fit the data to a coherent orbit.

We also carried out a line bisector analysis to enable us to
ascertain whether the RV variations may be attributed to
starspots.

 The CCF was computed for regions which include the
photospheric lines
 commonly used in the Doppler Imaging technique, while excluding
chromospheric and telluric lines (\citeauthor{queloz01}
\citeyear{queloz01}). We computed the bisector and, to quantify
the changes
 in the CCF bisector shape, also the bisector inverse slope (BIS).
 The BIS was defined as the difference of the average values of the top and bottom zones
 (we avoided wings and core of the CCF profile, due to errors
 of bisectors measurements, which are large in these zones). We studied the bisector variations only for FOCES04 run,
 as it was the more suitable data for the study. Three cases are
reported in bibliography (see for example \citet{queloz01}, \citet{aap},
 \citet{bonfils07}, etc.): a) anticorrelation, which
indicates that the RV variations are due to stellar activity (by
active regions at the stellar surface like spots or plages), b)
lack of correlation, which indicates the Doppler reflex motion
around the center of mass due to other bodies orbiting the star,
c) correlation, which, as pointed out by Martinez-Fiorenzano et
al. (2005), indicates that the RV variations are due to light
contamination from an unseen stellar companion.

\begin{figure}
\includegraphics [width=5.5cm,angle=270]{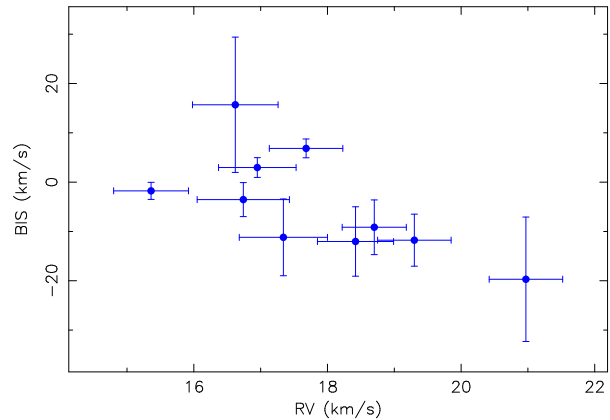}
 \caption{Bisector velocity span vs. RV for FOCES04 observing run.  The clear negative
 correlation indicates that RV variations are due to
 stellar activity (see Sect. \ref{52}).}
 \label{fig:bis}
\end{figure}

  As shown in Fig.~\ref{fig:bis}, there is an anticorrelation between BIS and RV, with a
Pearson correlation coefficient ($r$) of -0.6851. This result
suggests
  that the RV variations of FR~Cnc are due to stellar activity variations
 (e.g. spots on photosphere) and not due to a binary nature.
 When the spectrum of the standard star
 was broadened to the same rotational velocity of FR~Cnc, the profiles of the CCF became
 smoother and could be fitted with a Gaussian profile, (see Fig.~\ref{fig:perfil}, bottom).

 The irregular profiles of the CCF
(double peaks and asymmetries) can produce significant errors in
RV measurements (see Fig. \ref{fig:perfil}). These irregularities
may come from photospheric activity features on the stellar
surface. They can distort the profile of the photospheric lines
and induce variations in the peak of the CCF. However, this
behavior may be caused by the difference in rotational velocity
($v$~sin~$i$) between the target and standard star (see e.g.
\citeauthor{Galvez07} \citeyear{Galvez07}) when the standard is a
much slower rotator than the target. The CCF is essentially the
broadening function that would be applied
  to the template spectrum. Use of a broadened template removes higher
  moments introduced from the starspots, enabling a relatively unbiased estimate
 of RV to be determined (Fig.~\ref{fig:perfil}).
  A mean RV of 17.8 $\pm$ 1.6 is obtained for the 2004 data set
 and is in good agreement with the RV derived through optimisation
  of parameters in the Doppler imaging process (Section 7), where we find $V_{\rm hel}$ = 18.6 $\pm$ 0.6.
 Therefore all the radial velocities given in this paper have been
 calculated by cross-correlation with this rotational broadened spectrum of
 the standard star.

In Table~\ref{tab:vr} we list, for each spectrum, the heliocentric
radial velocities ($V_{\rm hel}$) with their corresponding errors
($\sigma_{V}$) obtained as weighted means of individual values
deduced for each
 order in the spectra. We also list data points from \citet{intr10} for comparison.

Those orders which contain chromospheric features and prominent
telluric lines have been excluded when determining the mean radial
velocity.

\subsection[]{Rotational Velocity} \label{43}

By using the program {\sc jstarmod} (see Sect. \ref{51spectral})
we have obtained the best fits for each observing run using $v
\sin{i}$ values of $\approx$35 km~s$^{-1}$. However, this value
depends on the rotation of the standard star which has non-zero
rotation. Therefore the obtained value can only be used as an
approximation.

To determine an accurate rotational velocity of this star we made
use of the following method (see \citeauthor{mart} \citeyear{mart}
for details). Rotational velocities, $v \sin{i}$ can be written as
follows (see \citet{Queloz98} and references therein):
\begin{equation}
\small{ \sigma_{rot}^2=\sigma_{obs}^2-\sigma_0^2  \Longrightarrow
v \sin{i}=A\sqrt{\sigma_{obs}^2-\sigma_0^2} } \label{eqRV}
\end{equation} where $A$ is a coupling constant which depends on the spectrograph
and its configuration. The spectrum of each of these stars was
broadened using the program  {\sc jstarmod}
 from $v$ sin$i$ = 1 km~s$^{-1}$ up to  50 km~s$^{-1}$  and the respective
 CCF was calculated. $A$ was found for every spectrograph by fitting
 the relation  $(v \sin{i})^2$ vs $\sigma_{obs}^{2}$.
 It is well known that $\sigma_{0}$ is a function
 of the broadening mechanisms which are present in the atmosphere of the
 star, except rotation (\citeauthor{Melo} \citeyear{Melo}). Since the broadening
 mechanisms are function of the temperature and gravity, we may expect
 a dependence of $\sigma_{0}$ on the temperature. To determine this
 dependence we use synthetic spectra with no rotational velocity computed using the {\sc atlas9} code by (\citeauthor{Kurucz} \citeyear{Kurucz}) adapted to work
 under a Linux platform by \citet{Sbordone04} and
 \citet{Sbordone05}. Once $A$ is determined and $\sigma_{0}$ calibrated with
 the color index ($B-V$), $\sigma_{obs}$ (width of the CCF of the star
 when is correlated with itself) is measured for each star,
 $v$sin$i$ can be directly  calculated using the above formula (\ref{eqRV}).

 In Table~\ref{tab:vr} we list, for each observing run, the averaged $v \sin{i}$
 value obtained. From Table~\ref{tab:vr} we estimate uncertainties of 1.9\,-3.2 km~s$^{-1}$ based on the standard deviations as each measurement epoch. It is likely that the $v$\,sin\,$i$ values vary by more than these uncertainties since FR~Cnc is very active and exhibits starspots (see Section 7) that significantly distort the rotationally broadened absorption lines. Moreover, while the observations taken in the year 2004 cover a complete rotation cycle, those $v$\,sin\,$i$ measurements at other epochs, only include one to two observations and are likely to yield more biased results (i.e. depending on the location of starspots at the observation phases). In Section 7, we model the line profile using our Doppler imaging code, allowing for the presence of spots. The resulting fits are thus likely to give a more accurate rotation velocity, $v$\,sin\,$i$, for FR~Cnc.

\begin{table}
\caption[Radial and Rotational Velocities] {Radial and Rotational
Velocities \label{tab:vr}}
\begin{center}
\scriptsize
\begin{tabular}{llcccccll}
\noalign{\smallskip} \hline \noalign{\smallskip}
 Run & HJD & \multicolumn{1}{c}{V$_{\rm hel}$ $\pm$ $\sigma_{\rm V}$} & $\overline{{\it v}sin{\it i}}$ \\
\noalign{\smallskip}
  & (2400000 +) &  \scriptsize (km~s$^{-1}$) & \scriptsize (km~s$^{-1}$)  \\
\noalign{\smallskip} \scriptsize
\\
\noalign{\smallskip} \hline \noalign{\smallskip}
FOCES04 & 53098.3713 & 16.95 $\pm$ 0.58 & 44.1 $\pm$ 1.9\\
FOCES04 & 53099.3411 & 18.42 $\pm$ 0.57 & \\
FOCES04 & 53099.4230 & 17.34 $\pm$ 0.66 & \\
FOCES04 & 53099.4546 & 16.74 $\pm$ 0.69 & \\
FOCES04 & 53100.3156 & 16.62 $\pm$ 0.64 & \\
FOCES04 & 53100.3917 & 15.36 $\pm$ 0.56 & \\
FOCES04 & 53101.3264 & 18.70 $\pm$ 0.48 & \\
FOCES04 & 53101.4495 & 20.97 $\pm$ 0.55 & \\
FOCES04 & 53102.3306 & 19.30 $\pm$ 0.55 & \\
FOCES04 & 53102.4753 & 17.68 $\pm$ 0.55 & \\
FOCES06 & 54086.6770 & 19.85 $\pm$ 0.63 & 37.2 $\pm$ -\\
FOCES06 & 54088.5918 & 19.62 $\pm$ 0.45 & \\
FOCES06 & 54091.6192 & 20.45 $\pm$ 0.53 & \\
FOCES07a & 54156.4482 & 19.98 $\pm$ 0.52 & 41.3 $\pm$ 2.7 \\
FOCES07a & 54158.5625 & 20.41 $\pm$ 0.80 & \\
FOCES07b & 54228.3461 & - & 43.6 $\pm$ 3.2\\
FOCES07b & 54229.3496 & - & \\
FIES08  & 54547.4408 & 17.82 $\pm$ 0.74 & -\\
Upgren02  & 51626.695 & 27 $\pm$ 2.3 & - \\
Upgren02  & 51626.730 & 24 $\pm$ 4.1 & - \\

\noalign{\smallskip} \hline
\end{tabular}
\end{center}
\end{table}

\subsection[]{Kinematics}
We computed the galactic space-velocity components ($U$, $V$, $W$)
and their associated errors of FR~Cnc using  the procedure
described by \citet{JohnsonSoderblom} modified by
 L\'opez-Santiago (see \citeauthor{Montes01a} \citeyear{Montes01a}). This procedure uses
J2000 coordinates and takes into account correlation in the
measures of \emph{Hipparcos}. We use averaged RV calculated here
(18.63 $\pm$ 0.14 km~s$^{-1}$)
 and the proper motions and parallax from \emph{Hipparcos}.

The obtained values of the components with its module $V_{\rm
Total}$ and associated errors are given in
 Table~\ref{tab:par2}. The velocity components in the  ($U$, $V$) diagram
are clearly within the young disc population boundaries (Eggen
1984a,b, 1989; Montes et~al. 2001a,b) indicating that the star
belongs to the young disc and that it might also belong to the
 IC 2391 moving group, mentioned previously by Pandey et~al. (2005), but
 the Eggen kinematic criteria (see \citet{Montes01a} for details)
 are negative,
 showing that FR~Cnc could be not a member of any moving group (MG hereafter)
in the young disc area.

 The classical view of MGs (e.g. Eggen 1984a), i.e. they come from the remnant
 of a star-forming cloud has been discussed in recent years.
 Several studies
 (e.g. Famaey et~al. 2007, 2008; Antoja et~al. 2008; Zhao et~al. 2009)
 seem to support a dynamic or resonant mechanism origin. While both theories are feasible, we
 will just take into account that the percentage of contamination of the young disc space velocity area
 by old field population is high
 (see \citeauthor{lopez09} \citeyear{lopez09} and reference therein) and so
 age constraints are needed to assess if FR~Cnc belongs to any of young disc moving groups.

\subsection{The Li~{\sc i} $\lambda$6707.8 line}
Li~{\sc i} $\lambda$6707.8 spectroscopic feature is an important
diagnostic tool for assessing the age in late-type stars,
 since Lithium is destroyed easily by thermonuclear reactions
in the stellar interior.

The spectral region of the resonance doublet of Li~{\sc i} at
$\lambda$6708 \AA\ is covered in all the high resolution
observations. We measured the equivalent width ($EW$ hereafter) in
seven
  spectra.

\begin{figure}
\includegraphics [width=7cm]{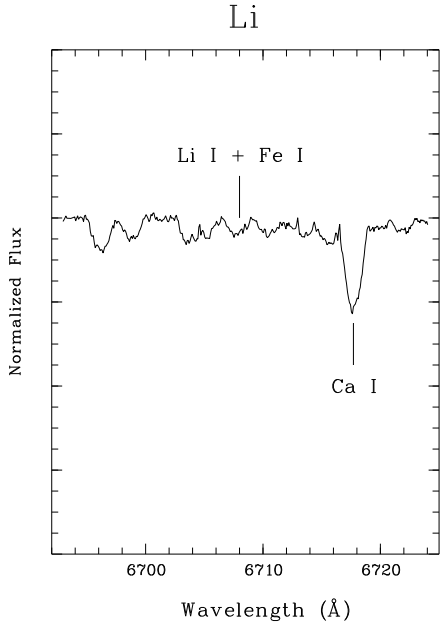}
 \caption{Spectrum in the Li~{\sc i} line region, resulting of co-adding the 10
observed spectra of FOCES04 run in order to increase the S/N. The
position of the  Li~{\sc i} +  Fe~{\sc i} and  Ca~{\sc i} are
marked.}
 \label{fig:li}
\end{figure}

Due to the small value, we were not able to measure $EW$s
directly,
 or use spectral subtraction technique. Therefore $EW$s have been obtained
 using the IRAF task { \sc sbands}, performing an integration within a band of
 1.6 \AA \ centred in the lithium line (Maldonado et al. 2010).
  We have obtained an averaged value of 54 m\AA. In our high resolution spectra the Li~{\sc i} line
 is blended with Fe~{\sc i} 6707.4 \AA. To subtract the Fe~{\sc i} contribution we used
 the color-index relation from \citeauthor{Favata} (\citeyear{Favata}).
 Therefore, we obtain the final value of $EW$(Li~{\sc i}) to be 34 m\AA.
 In Fig.~\ref{fig:li} we plot as an example, a spectra formed by
  co-adding the FOCES04 run spectra and we indicated the position of the Li~{\sc i} line.

 By comparing this value with stars of similar spectral type (K5-K7 with V-I = 1.3) in
 other MGs members or clusters of a well-known age (in the same way as
 in Figure 3 of
 \citeauthor{lopez09} \citeyear{lopez09}),
  it is in agreement with being a young object between
  10-120 and compatible with being IC 2391 MG member.

\subsection{Other age indicators}

Ideally, further constraints on FR~Cnc's age should be calculated
to confirm its youth. Popular age approximations are for example
the X-ray flux-age relation
 (\citeauthor{Mamajek} \citeyear{Mamajek}; equation A3) or the commonly used relation between age and $R^{'}_{HK}$ index, that measures chromospheric emission
 in the cores of the broad chromospheric Ca~{\sc ii} H \& K lines
 (see e.g. \citeauthor{Noyes} \citeyear{Noyes};  \citeauthor{Baliunas} \citeyear{Baliunas};
 \citeauthor{Mamajek} \citeyear{Mamajek}).
 The former relation is valid for stars with spectral types earlier than that of FR~Cnc
 and so can not be used.
 Also, using the latter relation, from our Ca~{\sc ii} H \& K fluxes (see Sect.~\ref{63} and
 Table~\ref{tab:vr}) we obtain a log$R^{'}_{HK}$  = -3.48 which is outside the validity range
 of the activity-age relation but in this case it is
 compatible with the young age.

\section[]{Chromospheric activity indicators}
Both echelle and long slit spectra analysed in this work allowed
us to study the behavior of the different indicators from the
Ca~{\sc ii} H \& K to the Ca~{\sc ii} IRT lines, which are formed
at different atmospheric altitudes. The chromospheric contribution
to these features was determined by using the spectral subtraction
technique described in detail by \citet{Montes00b} and
\citet{Galvez02}.

The excess emission $EW$ of different spectral features were
measured in the subtracted spectra. In Table \ref{tab:ew} we give
the $EW$ for the Ca~{\sc ii} H \& K, H$\epsilon$, H$\delta$,
H$\gamma$, H$\beta$, H$\alpha$, and Ca~{\sc ii} IRT
($\lambda$$\lambda$8498, 8542, 8662 \AA)
 lines for the echelle spectra.
These $EW$s were converted to an absolute surface fluxes by using
the empirical stellar flux scales calibrated by \citet{Hall} as a
function of the star color index. In our case, we used the $B-V$
index and the corresponding coefficients for Ca~{\sc ii} H \& K,
H$\alpha$ and Ca~{\sc ii} IRT, using for H$\epsilon$ the same
coefficients as for Ca~{\sc ii} H \& K, and derived the H$\delta$,
 H$\gamma$ and H$\beta$ coefficients of flux by carrying out an interpolation
 between the values of Ca~{\sc ii} H \& K and H$\alpha$.
 The logarithm of the obtained absolute flux at the stellar surface
 (log$F$$_{\rm S}$) in ergs~cm$^{-2}$~s$^{-1}$~\AA$^{-1}$ for the
 different chromospheric activity indicators is given in Table \ref{tab:fl}.

\begin{figure}
\includegraphics [width=8cm]{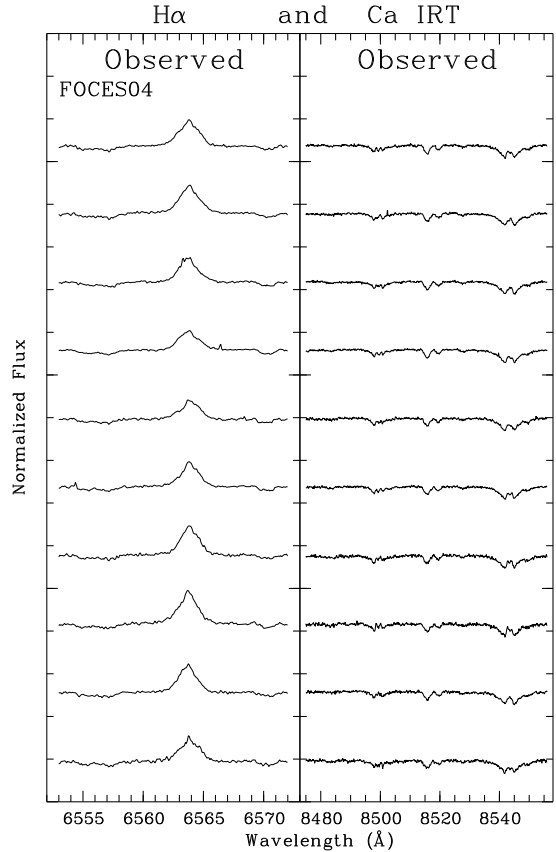}
 \caption{Spectra in the H$\alpha$ (left side)
 and Ca~{\sc ii} IRT $\lambda\lambda$8498, 8542 (right side) line regions
 for FOCES04 observing run;
 clear wide and prominent emission arises over the continuum from H$\alpha$
 and a clear emission in the core of the absorption line is seen in
 the Ca~{\sc ii} IRT lines.
 }
 \label{fig:hairt}
\end{figure}

\begin{figure}
\includegraphics [width=6.5cm,angle=270]{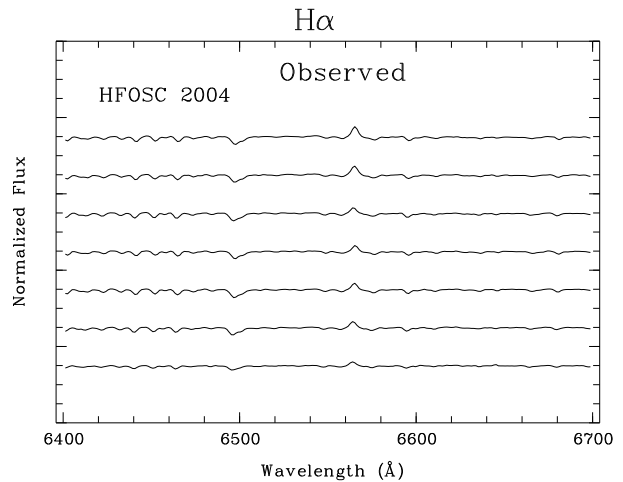}
 \caption{Sample of spectra in the H$\alpha$ line region
 for HFOSC04 observing run; clear emission arises over the continuum from H$\alpha$
 in this low resolution spectra.
 }
 \label{fig:haindia}
\end{figure}

\begin{figure}
\includegraphics [width=6.0cm,angle=270]{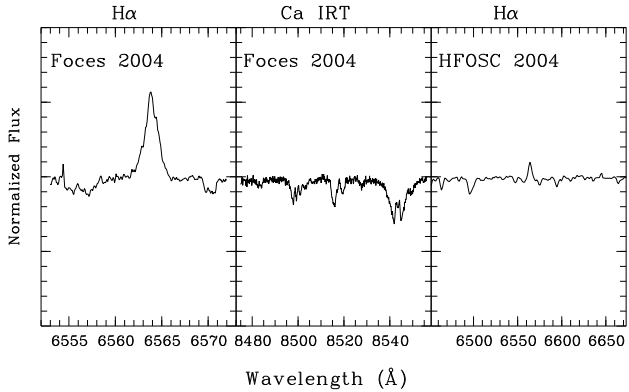}
 \caption{A representative spectra of Figs.~\ref{fig:hairt} and ~\ref{fig:haindia},
 showing a zoom of the H$\alpha$ (left side) and Ca~{\sc ii} IRT (center) line regions
 for FOCES04 and H$\alpha$ (right) line region for HFOSC04.}

 \label{fig:hairtzoom}
\end{figure}

Fig.~\ref{fig:hairt} shows representative observations in the
H$\alpha$,
 and Ca~{\sc ii} IRT $\lambda$$\lambda$$8498$, 8542 line regions for
 high resolution spectra.
Fig.~\ref{fig:haindia} shows representative observations in the
H$\alpha$
 for low resolution spectra. Fig.~\ref{fig:hairtzoom} shows a closer view of one spectrum from Figs. ~\ref{fig:hairt}
 and ~\ref{fig:haindia} where the emission can be better seen.

\subsection{The H$\alpha$ line} \label{Ha}

We analysed the H$\alpha$ line region for all the spectra.
 This line in the obtained spectra is always observed in emission above the continuum (see Figs.~\ref{fig:hairt} and ~\ref{fig:haindia}).

 Measuring the $EW$ of this line, we found that the
 $EW$ average of the H$\alpha$ emission is quite
 different in every season, showing significant variability in time-scales of a year.
 $EW$(H$\alpha$)~= 3.23  \AA\ for FOCES04 run while,
 $EW$(H$\alpha$)~= 1.67, 1.72 and 1.87  \AA\ for FOCES06,
 FOCES07a and FOCES07b respectively, $EW$(H$\alpha$)~= 1.72
 for the only value of FIES08 run. For the low resolution
 spectra we have an average value (in 37 spectra taken during
 three consecutive nights) of $EW$(H$\alpha$)~=~2.14~\AA\ in HFOSC04 run.

\begin{figure}
\includegraphics [width=3.3cm,angle=270]{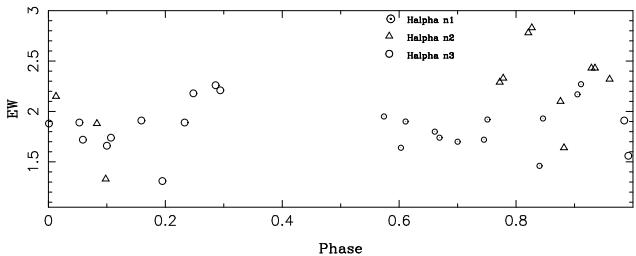}
 \caption{Variation of H$\alpha$ $EW$ vs phase in HFOSC04 run.
 Different symbols represent different nights. The
 second night is in triangles, showing higher values of the  H$\alpha$ $EW$s.}
 \label{fig:haindiam11}
\end{figure}

\begin{figure}
\includegraphics [width=7.3cm,angle=270]{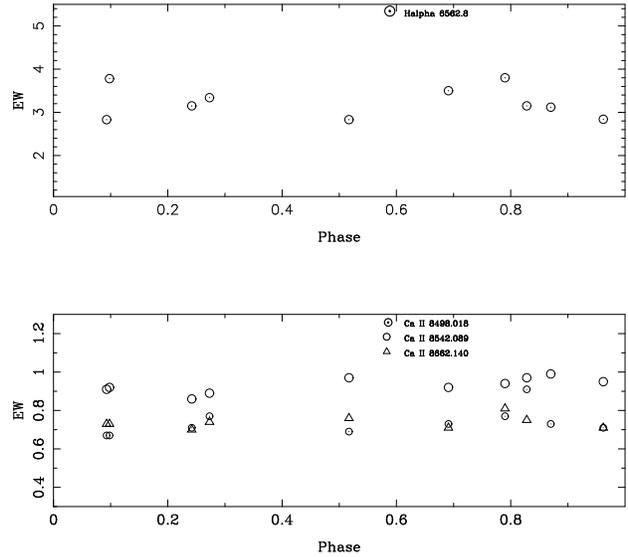}
 \caption{Top: Variation of H$\alpha$ $EW$ vs phase in FOCES04 run.
 Bottom: Variation of Ca~{\sc ii} IRT $EW$ vs phase.}
 \label{fig:hairtfec11}
\end{figure}

Fig. ~\ref{fig:haindiam11} shows the variation of $EW$ vs phase (calculated with
 the photometric period) in HFOSC04 run. Different symbols
 represent different nights. The
 second night (triangles) shows higher values of the  H$\alpha$ $EW$s.
 Top of Fig. ~\ref{fig:hairtfec11}, represent the  variation of $EW$ vs phase
 in the FOCES04 run.

 Comparing the variations of H$\alpha$ $EW$ between the runs and the variation
 in each run when possible, we note activity level variations on a month-long timescale,
 from one year to the next (see Table \ref{tab:ew}). Making the comparison with the photometry, we notice a correspondence between
 marked variations in the light curve and $EW$(H$\alpha$).
 In the year 2004 both photometry and spectroscopy show
  a high level of FR~Cnc activity, while it is
 decreasing rapidly in the year 2005 and then remains on that level during our further observations. This can be interpreted as an
 activity cycle of at least 4--5 years, similar to Sun or other stars activity cycles,
 but further follow up is needed to confirm this. Accurate stellar activity cycle
 can prove useful for study the dynamo interface and activity
cycle-rotation-spectral type
 mechanisms in the stars (see e.g. Lorente \& Montesinos 2005).

 The persistence of H$\alpha$ emission indicates that it is a very active BY Dra system, but the vast range of
 variability levels make this star unusual and interesting for further study.

\subsection{The H$\beta$, H$\gamma$ and H$\delta$ lines}

\begin{figure}
\includegraphics [width=6cm,angle=270]{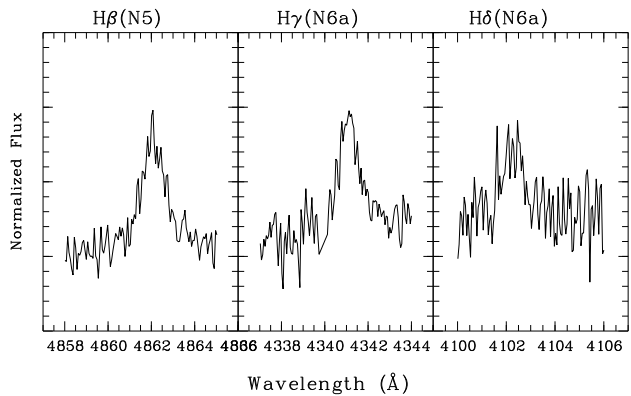}
 \caption{Representative subtracted spectra of H$\beta$, H$\gamma$ and H$\delta$
  lines in different nights of FOCES04 run.
 }
 \label{fig:hbdg}
\end{figure}

We can see the absorption of H$\beta$, H$\gamma$ and H$\delta$
Balmer
 lines filled in with emission in the observed spectra.
 Fig.~\ref{fig:hbdg} plots a representative subtracted spectra of these three lines in different
 nights of FOCES04 run.

 The variation of these lines with rotational phase and from season to season follows the same
 trend as H$\alpha$  variation.

 We also measured the ratio of excess emission in the H$\alpha$ and H$\beta$ lines
 $(\frac{EW({\rm H\alpha})}{EW({\rm H\beta})})$ and the ratio
 of excess emission $\frac{E_{\rm H\alpha}}{E_{\rm H\beta}}$
 with the correction: \begin{equation}
\frac{E_{\rm H\alpha}}{E_{\rm H\beta}} =
 \frac{EW({\rm H\alpha})}{EW({\rm H\beta})}\cdot0.2444\cdot2.512^{(B-R)}
\end{equation} given by \citet{HallRamsey}. This corrects the absolute flux density in these lines for the color difference
 in the components. We have obtained a mean value of
 $\frac{E_{\rm H\alpha}}{E_{\rm H\beta}}$~$\approx2.5$. This
 value is in the limit between the presence of prominence-like material on or above the stellar
 surface (\citeauthor{Buz89} \citeyear{Buz89} and \citeauthor{HallRamsey} \citeyear{HallRamsey}).

\subsection{Ca~{\sc ii} H \& K and H$\epsilon$}\label{63}

The Ca~{\sc ii} H \& K line region is included in most of the spectra
 but the efficiency of the spectrograph and the CCD decreases very
 rapidly due to the position of these lines at the end of the echellogram. Therefore, the obtained S/N ratio is very low,
 and the normalization of the spectra is very difficult.
 In many cases we could not measure the $EW$ lines and in other we
 measured them in the observed spectra as it was not possible
 to apply the spectral subtraction for this region (see Tables \ref{tab:ew} and \ref{tab:fl}).

 Strong emission in the Ca~{\sc ii} H \& K is seen
 despite the low S/N.

\subsection{Ca~{\sc ii} IRT lines
($\lambda\lambda$8498, 8542, and 8662)}

The three lines of the Ca~{\sc ii} (IRT) are included in all our
echelle spectra. In all of them a clear emission arising in the
core of the absorption lines is seen (see Fig.~\ref{fig:hairt}).

 Averaged values of emission $EW$ are $EW$(Ca~{\sc ii})~= 0.74, 0.93 and 0.81 \AA\ for
 $\lambda\lambda$8498, 8542, and 8662 in FOCES04 run.
 $EW$(Ca~{\sc ii})~= 0.49,  0.72 and 0.61 \AA\ in FOCES06 run.
 $EW$(Ca~{\sc ii})~= 0.66, 0.86 and 0.70 \AA\ in FOCES07a.
 $EW$(Ca~{\sc ii})~= 0.79, 1.03 and 0.77 \AA\ in FOCES07b and
 $EW$(Ca~{\sc ii})~= 0.71, 0.89 and 0.79 \AA\ in FIES08.

 The variation of the emission in these lines is significant although not
 as strong as in H$\alpha$ (see Table \ref{tab:ew}).

 Only for the FOCES04-run  we have enough data points to
 compare variations between H$\alpha$ $EW$s and Ca~{\sc ii} $EW$s.
 In other active stars a clear anticorrelation is usually seen
 (as their emissions come from different features in the stellar surface), see e.g. stars in \citet{arevalo99},
 \citet{Montes00b}, \citet{Galvez02}, \citet{Galvez09} etc.,
 but it is very weak in the case of FR~Cnc (see Fig. ~\ref{fig:hairtfec11}).

In addition, we have calculated the ratio of excess emission $EW$,
$\frac{E_{8542}}{E_{8498}}$, which is also an indicator of the
type of chromospheric structure, which produces the observed
emission. In solar plage values of $\frac{E_{8542}}{E_{8498}}$
$\approx$~1.5--3 are measured, while in solar prominence the
values are $\approx$~9, the limit of an optically thin emitting
plasma (\citeauthor{chester} \citeyear{chester}). We have found a
$\frac{E_{8542}}{E_{8498}}$$\approx$1.3,
 indicating that Ca~{\sc ii} IRT emission comes from plage-like
 regions.

\section{Doppler imaging}

Since FR~Cnc is a rapid rotator with considerable broadening of
spectral lines, we generated an indirect starspot map using the
Doppler Tomography of Stars (DoTS)
 imaging code (\citeauthor{cameron01mapping} \citeyear{cameron01mapping}). In order to detect the line distortion due
 to starspots in the high resolution spectra, we have applied least-squares deconvolution
 (\citeauthor{donati97zdi} \citeyear{donati97zdi}; \citeauthor{barnes98aper} \citeyear{barnes98aper}) to the 4362--6845 \AA ~wavelength region of FOCES04 spectra (obtained at 2004 March -- April).
 A single line, free of the effects of rotational line blending and with high S/N is thus
 derived. Deconvolution is carried out using a $T$ = 2450 K model line list (VALD; see \citeauthor{Kupka1} \citeyear{Kupka2} \& \citeyear{Kupka1}) which
 indicates that there are 8345 lines in the selected wavelength region with normalized depths
 of 0.05--1.0. Regions around hydrogen Balmer lines, the Mg triplet and Na doublet are
 excluded from the deconvolution. A single line profile with a mean of
 $\lambda$ = 5460.4 \AA \ is derived for each observed spectrum. The mean S/N of the input spectra over the entire 4362\AA\ - 6845\AA\ wavelength region was 21.8, while the mean
deconvolved line profiles possess S/N = 1008 (indicating a gain of
46.2). It should be noted that while the profile shown in
Fig.~\ref{fig:perfil} (upper panel) represents the
  broadening function of FR~Cnc (i.e. rotational velocity plus starspot distortions), the LSD profiles in Section
 7 are deconvolved using a linelist rather than a template. As such, the LSD profiles still contain the intrinsic
 stellar profile, and any distortions due to starspots will appear to possess a lower amplitude when compared with a template-derived CCF.

\begin{figure}
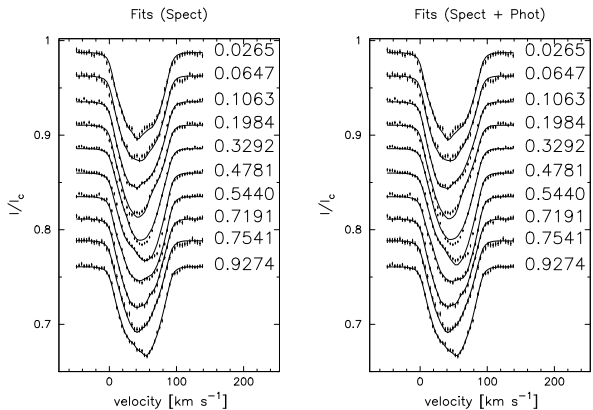

\includegraphics[width=60mm,angle=270]{s_frcnc_spect.ps}
\includegraphics[width=60mm,angle=270]{s_frcnc_spect_phot.ps}
\caption{The figure shows the deconvolved profiles, with the
phases of observation.
 The variation of the profile shape, due to starspots rotating into and out of view, is clearly seen.}
 \label{fig:xxx}
\end{figure}

\begin{figure*}
\includegraphics [width=8cm,angle=270]{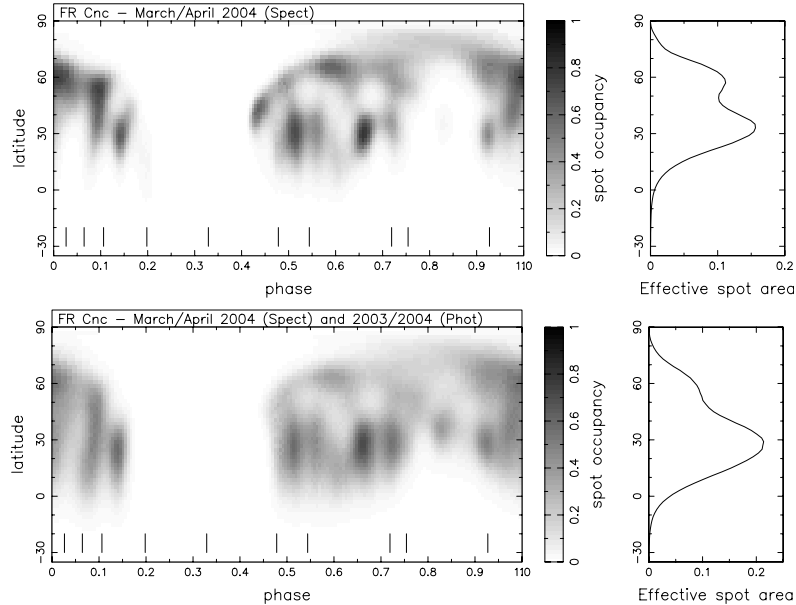}
 \caption{Mercator projection of the starspot image of FR~Cnc. A high degree of spot coverage is revealed.
 There are no visible spots in the phase range 0.1\,--\,0.3.
 FR~Cnc shows only mid-latitude starspots and
 no polar spot like other typical late K stars. The vertical lines mark the corresponding phase for 10 spectra we have used for deconvolution. The right hand
 plot shows how the mean effective spot area changes with
 latitude.}
 \label{fig:yyy}
\end{figure*}

For the imaging procedure, we used the standard star, HD 151877,
to
 represent the local intensity profile of a slowly rotating star. A two temperature
 model with $T_{\rm phot}$ = 4250 K and $T_{\rm spot}$ = 3000 K was used. The starspot
 image therefore represents the spot filling factor. Details of the Doppler imaging
 technique can be found in \citet{cameron01mapping}. We optimized the goodness of fit to the 10 deconvolved profiles
for heliocentric RV, axial inclination, equivalent width and
$v$\,sin\,$i$, finding $V_{hel}$ = 18.6 $\pm$ 0.6, $v$\,sin\,$i$ =
46.2 $\pm$ 0.8 km~s$^{-1}$ and i = 55 $\pm$ 5\degr. The
$v$\,sin\,$i$ value is higher than, but still consistent with, our
mean $v$\,sin\,$i$ value derived in Section 5.3. Since we here use
all rotation phases to derive our best fit Doppler image, it is
 likely that the result is less biased than the previous results in Section 5.3 that do not take account of the
  presence of starspots, which affect the profile shape. Starspot distortions in the wings of the profiles
  for instance, are likely to lead to underestimations of $v$\,sin\,$i$
 when using a single spectrum, leading to a systematically lower mean estimation of $v$\,sin\,$i$. We note however
 that our result for the 2004 observations is consistent with the value tabulated in Table 9 (i.e. 44.1 $\pm$ 1.9 km s$^{-1}$).  Fig.~\ref{fig:xxx} shows the deconvolved profiles (and phases of observation) while Fig.~\ref{fig:yyy} shows mercator projections of the starspot image of FR~Cnc.

\begin{figure}
\includegraphics [width=8cm]{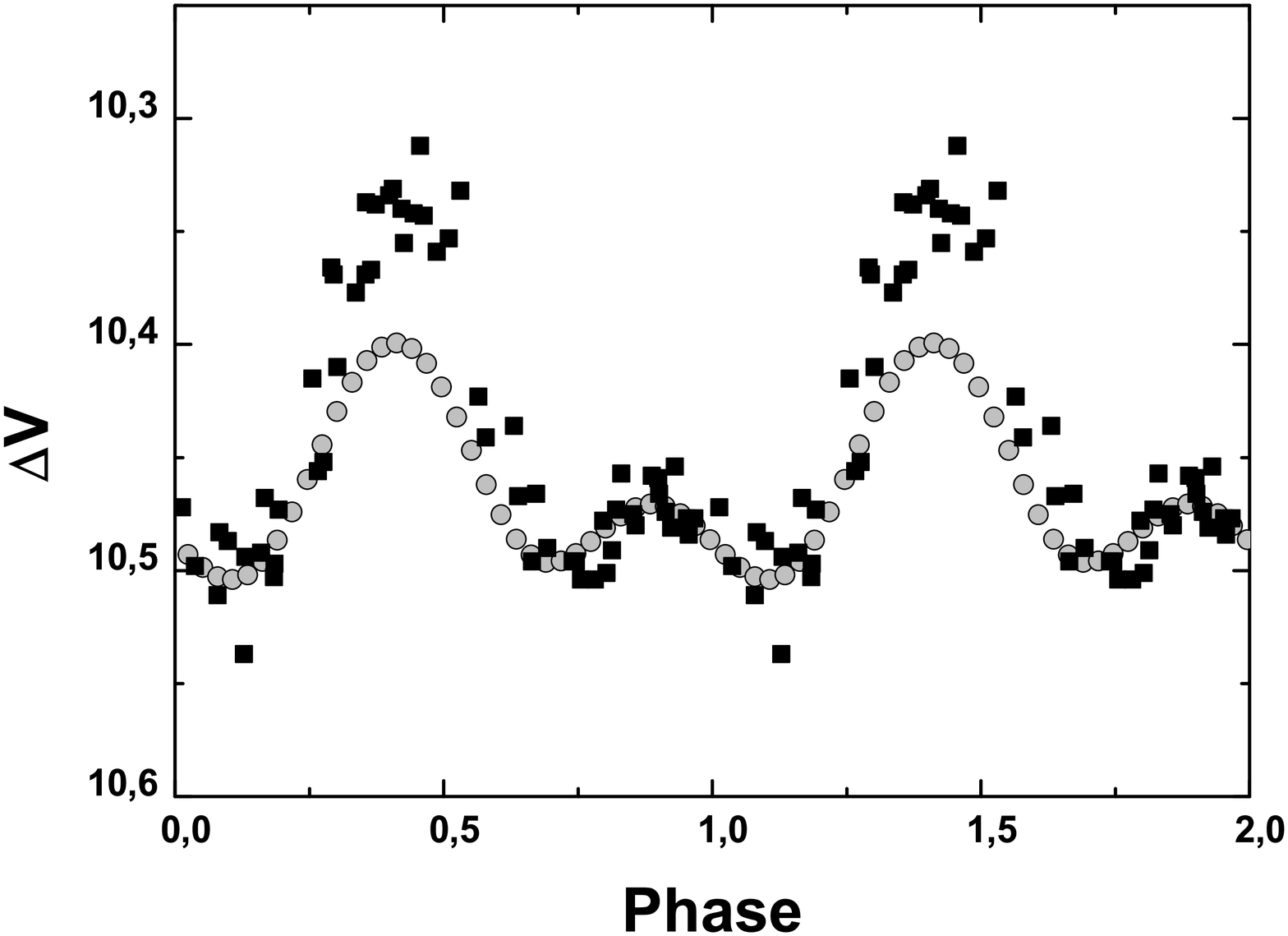}
\includegraphics [width=8cm]{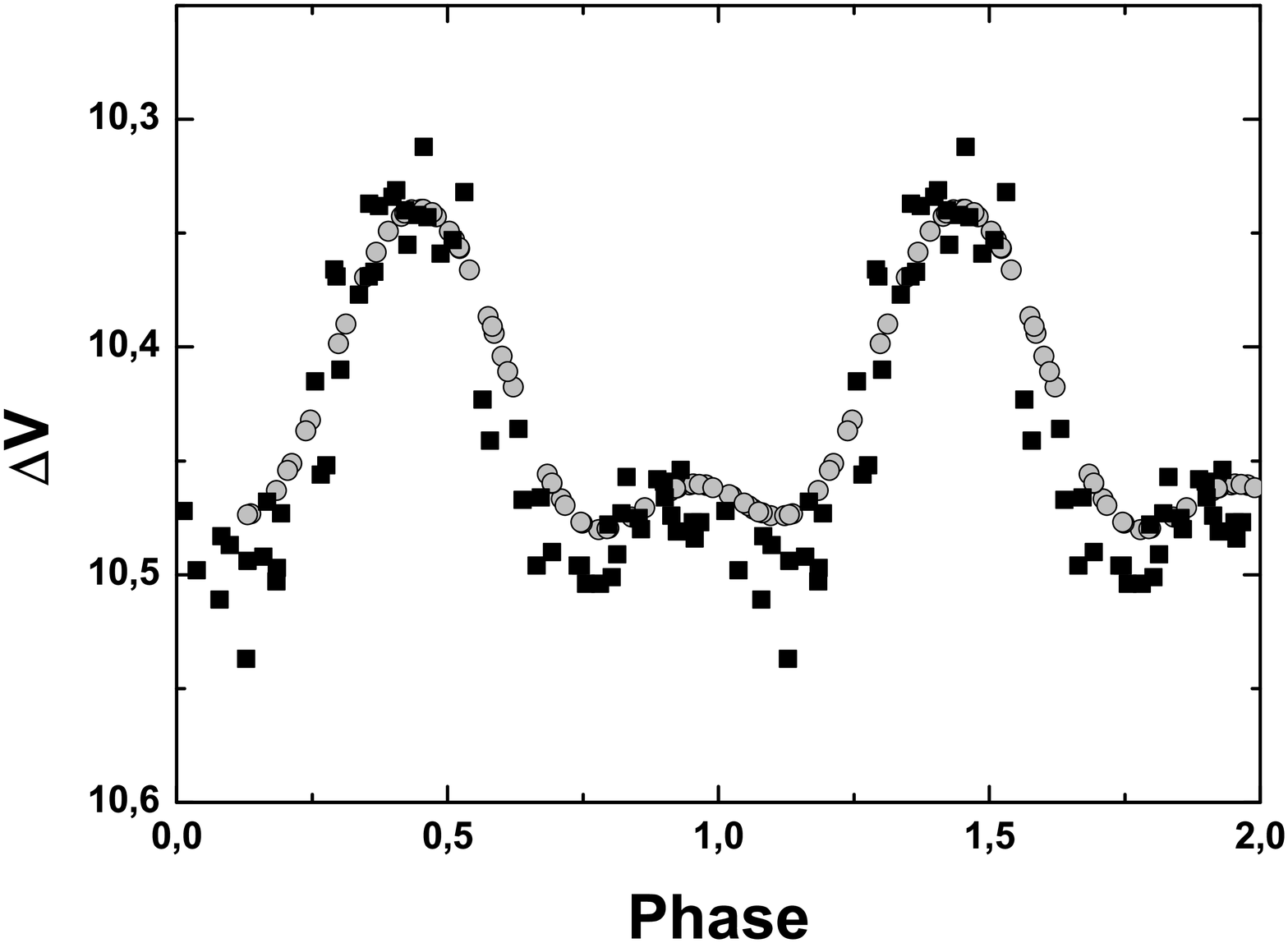}
 \caption{Photometric lightcurves calculated from the surface maps
and superimposed on \emph{ASAS-3}-lightcurve for corresponding
season of observations.}
 \label{fig:lighcid}
\end{figure}

The surface map in Fig.~\ref{fig:yyy} (upper panel) indicates that
FR~Cnc possesses considerable spot coverage as
 suggested by the time varying distortions in the deconvolved profiles in Fig.~\ref{fig:xxx}
(left panel - spectroscopic data only). For comparison with the
2003\,-\,2004 (JD = 2452934-2453146 in Table 1)
 lightcurve presented in Fig.~3 (top right), we have generated a synthetic lightcurve based on the spectroscopically
 derived image in Fig.~\ref{fig:yyy} (upper panel). While we are able to recover the observed lightcurve morphology,
  we considerably {\em underestimate} the amplitude (Fig.~\ref{fig:lighcid}, upper panel). This is likely due,
 at least in part, to loss of information in the Doppler imaging process (i.e. of the spectroscopic data) owing to
 resolution and noise constraints imposed by the fitting procedure which must optimise both the fit and the
 image entropy. With higher S/N ratio data and spectral resolution it is probable that more precise fitting
 would yield spot filling factors indicating cooler spots. In addition, there is a tendency for spots close to
 the equator ($<$20\degr) to be reconstructed at slightly higher latitudes than they are physically situated.
 This is due to the smaller entropy penalty of placing a smaller spot away from the equator as compared with
 a larger spot nearer the equator. The resulting spot distribution will naturally lead to predicted lightcurves
 with underestimated amplitudes since equatorial spots yield higher
 amplitudes.

We find a spot filling factor of 6 per cent from our spectroscopic
Doppler image. This is at the lower end of the typical values (up
to $\sim 30$ per cent) found for active stars by O'Neal, Saar \&
Neff (1996), O'Neal,
 Neff \& Saar (1998) and O'Neal et al. (2004) using TiO indicators. We have therefore reconstructed an image
 that also makes use of the 2003/2004 $V$-band photometric data (e.g. see Barnes et al. 1998). It should be
 noted that while the lightcurve data spans several months, the spectroscopic data were taken over only a few
 days. Since the photometric observations taken within $\pm~1$ month of the spectroscopic data do not show
 significant deviation from the photometric dataset as a whole, we made use of all photometric points.
 The resulting simultaneous fits to the data are shown in Fig.~\ref{fig:xxx} (right panel) and Fig.~\ref{fig:lighcid}
 (bottom panel), while the reconstructed image is shown in Fig.~\ref{fig:yyy} (bottom panel).
Simultaneous imaging that makes use of spectroscopic and
photometric data must strike a balance, such that reasonable fits
to both data sets are achieved. We have already outlined (above)
possible modest biases in an image derived from a spectroscopic
fit to finite S/N. Lightcurve data alone does not contain
sufficient information to make reliable images of single stars.
Any complex spot structure or spot groups will be rendered simply
as a single spot. Photometric imaging favours reconstruction at
lower latitudes because a smaller spot is more easily able to
reproduce the amplitude variation (and yields a lower image
entropy penalty) in an observed lightcurve than a slightly larger
spot at higher latitude. Allowing the photometric fit to more
closely match the observed lightcurve results in a slightly poorer
fit to the spectroscopic data. The fitted lightcurve now closely
matches the observed lightcurve, since more large scale (albeit
unresolved) starspot structure has been recovered. This has
necessitated some loss of image details (Fig.~\ref{fig:yyy}, lower
panel) as compared to the spectroscopic reconstruction alone
(Fig.~\ref{fig:yyy}, upper panel). The filling factor is now
  greater at 9 per cent, adding further weight to the argument that spectroscopy alone is unable to recover all
 starspot information. Indeed, if the results from TiO are to be believed, there may be further unresolved global starspot covereage on FR~Cnc and all other stars that have been
subjected to such studies.

We also carried out a bisector analysis on the LSD profiles,
calculating the bisectors in the same way as in Sect. 5.2.  As we
have mentioned, the LSD profiles should show a smoothing of the
bisectors, reflected in the lower value of the bisector we
obtained comparing with the original spectra. As expected, the
anticorrelation is still seen, but here we have a strongest
Pearson correlation coefficient of r = -0.8992.

 As might be expected for such an active star, FR~Cnc exhibits a high degree of spot coverage.
  No spots are visible in the phase range 0.1\,--\,0.3. This agrees well with the photometric
  $V$-band light curve plotted for the same period (Fig. \ref{asas_phase}, 2003--2004) which shows a maximum at
 phase $\sim 0.2$. This contrast with a significant degree of spotness at other longitudes
 is likely responsible for the high degree of photometrical modulation seen in this star.
 The starspot distribution of FR~Cnc is also of interest since it is one of the latest
 spectral types to have been imaged. Comparison with the K5V rapid rotator LO Peg, \citet{barnes05lopeg},
 shows a significantly different spot distribution. While LO Peg rotates more rapidly
 ($v$~sin~$i$ = 65.9 km~s$^{-1}$), showing predominantly mid-high latitude spot structures and a polar
 spot, FR~Cnc shows only mid-latitude starspots. Small scale spot structure is expected
 to be variable on a day-long timescale, while \citet{jeffers07abdor} have shown that the
  polar spot on the K0V star, AB Dor, is variable in extent over periods of years.
  While G, and K stars such as AB Dor often show strong polar caps, no such features were seen
 on the M1--M2 dwarf stars HK Aqr and RE 1816+541 (\citeauthor{barnes01mdwarfs} \citeyear{barnes01mdwarfs}).
 This change in
 starspot location (i.e. with latitude) may be due to a change from a convective shell-type
 dynamo to a fully convective dynamo. Predominantly mid-high latitude spots are expected
 to arise from a rapidly rotating solar-like dynamo (\citeauthor{schussler96buoy} \citeyear{schussler96buoy}; \citeauthor{granzer00} \citeyear{granzer00}) while
  a distributed dynamo may be expected to produce spots at all
latitudes. However, only further spectroscopic time series observations
of FR~Cnc, with a higher cadence of observations (enabling better
surface resolution to be obtained), would reveal the long term stability
of the starspot patterns.

\begin{table*}
\caption[$EW$ of Chromospheric Indicators] {$EW$ of Chromospheric
Indicators \label{tab:ew}}
\begin{center}
\scriptsize
\begin{tabular}{cccccccccccc}
\noalign{\smallskip} \hline \noalign{\smallskip}
      & \multicolumn{10}{c}{EW(\AA) in the subtracted spectra} \\
\cline{3-12} \noalign{\smallskip}
 Run & HJD & \multicolumn{2}{c}{CaII} & & & & & &
\multicolumn{3}{c}{CaII IRT} \\
\cline{3-4}\cline{10-12} \noalign{\smallskip} & (2400000+)   &
 K   & H  & H$\epsilon$ & H$\delta$ & H$\gamma$ & H$\beta$ & H$\alpha$ &
$\lambda$8498 & $\lambda$8542 & $\lambda$8662 \scriptsize
\\
\noalign{\smallskip} \hline \noalign{\smallskip}
FOCES04 & 53098.3713 & $^{\ddag}$ & $^{\ddag}$ & $^{\ddag}$ & $^{\ddag}$ & 0.42 & 1.13 & 2.83 & 0.69 & 0.97 & 0.76\\
FOCES04 & 53099.3411 & 2.46 & $^{\ddag}$ & $^{\ddag}$ & 0.62 & 1.10 & 1.32 & 3.50 & 0.73 & 0.92 & 0.71\\
FOCES04 & 53099.4230 & $^{\ddag}$ & $^{\ddag}$ & $^{\ddag}$ & $^{\ddag}$ & 0.40 & 1.44 & 3.80 & 0.77 & 0.94 & 0.81\\
FOCES04 & 53099.4546 & $^{\ddag}$ & $^{\ddag}$ & $^{\ddag}$ & $^{\ddag}$ & 0.46 & 1.10 & 3.15 & 0.91 & 0.97 & 0.75\\
FOCES04 & 53100.3156 & 1.54 & 2.74 & 1.70 & 0.21 & 0.67 & 1.30 & 3.12 & 0.73 & 0.99 & 2.18\\
FOCES04 & 53100.3917 & 1.58$^{\dagger}$ & 1.47 & 0.73 & 0.25 & 0.30 & 1.03 & 2.84 & 0.71 & 0.95 & 0.71\\
FOCES04 & 53101.3264 & $^{\ddag}$ & $^{\ddag}$ & $^{\ddag}$ & 0.40 & 0.76 & 1.08 & 2.83 & 0.67 & 0.91 & 0.73\\
FOCES04 & 53101.4495 & $^{\ddag}$ & $^{\ddag}$ & $^{\ddag}$ & $^{\ddag}$ & 0.27 & 1.23 & 3.15 & 0.71 & 0.86 & 0.70\\
FOCES04 & 53102.3306 & 2.12 & 1.74 & 1.02 & $^{\ddag}$ & 0.77 & 1.21 & 3.78 & 0.67 & 0.92 & 0.73\\
FOCES04 & 53102.4753 & $^{\ddag}$ & $^{\ddag}$ & $^{\ddag}$ & $^{\ddag}$ & $^{\ddag}$ & 1.40 & 3.34 & 0.77 & 0.89 & 0.74\\
FOCES06 & 54086.6770 & 6.58$^{\blacktriangle}$ & 6.82$^{\blacktriangle}$ & 0.00$^{\blacktriangle}$ &  0.36 & 0.30 & 0.74 & 1.49 & 0.29 & 0.63 & 0.58 \\
FOCES06 & 54088.5918 & 3.72$^{\blacktriangle}$ & 4.30$^{\blacktriangle}$ & 1.00$^{\blacktriangle}$ & 0.26 & 0.73 & 0.50 & 1.93 & 0.59 & 0.53 & 0.66    \\
FOCES06 & 54091.6192 & 5.84$^{\blacktriangle}$ & 5.39$^{\blacktriangle}$ & 0.90$^{\blacktriangle}$ & 0.37 & 0.36 & 0.58 & 1.60 & 0.60 & 0.99 & 0.58 \\
FOCES07a & 54156.4482 & 9.69$^{\blacktriangle}$ & 5.97$^{\blacktriangle}$ & 1.40$^{\blacktriangle}$ & 0.32 & 0.24 & 0.74 & 1.80 & 0.68 & 0.86 & 0.71  \\
FOCES07a & 54158.5625 & 9.58$^{\blacktriangle}$ & 3.78$^{\blacktriangle}$ & 0.48$^{\blacktriangle}$ & - & 0.13 & 0.76 & 1.65 & 0.64 & 0.86 & 0.69 \\
FOCES07b & 54228.3461 & - &  -  &  - & - &  - & - & 1.87 & 0.79 & 1.03 &  0.77 \\
FOCES07b & 54229.3496 & -  &  -   &   - &  - &  -  & -  & 1.72 & 0.71 & 0.89 &  0.79    \\
\noalign{\smallskip} \hline
\end{tabular}
\\
{\scriptsize
$^{\dagger}$ Values measured with low S/N.\\
$^{\ddag}$ Values not measured due to the low S/N.\\
$^{\blacktriangle}$ Values measured in the observed spectra (spectral subtraction is not applied).\\
}
\end{center}
\end{table*}
\begin{table*}
\caption[Emission Fluxes] {Emission Fluxes \label{tab:fl}}
\begin{flushleft}
\scriptsize
\begin{center}
\begin{tabular}{cccccccccccc}
\noalign{\smallskip} \hline \noalign{\smallskip}
           \multicolumn{12}{c}{~logF$_{\rm S}$} \\
\cline{3-12} \noalign{\smallskip} Run & HJD &
\multicolumn{2}{c}{CaII} & & & & & &
\multicolumn{3}{c}{CaII IRT} \\
\cline{3-4}\cline{ 10-12} \noalign{\smallskip} & (2400000+)     &
 K   & H  & H$\epsilon$ & H$\delta$ & H$\gamma$ & H$\beta$ & H$\alpha$ &
\scriptsize $\lambda$8498 & \scriptsize $\lambda$8542 &
\scriptsize $\lambda$8662
\\
\noalign{\smallskip} \hline \noalign{\smallskip}
FOCES04 & 53098.3713 & $^{\ddag}$ & $^{\ddag}$ & $^{\ddag}$ & $^{\ddag}$ & 5.88 & 6.35 & 6.91 & 6.24 & 6.38 & 6.28 \\
FOCES04 & 53099.3411 & 6.62 & $^{\ddag}$ & $^{\ddag}$ & 6.03 & 6.30 & 6.42 & 7.00 & 6.26 & 6.36 & 6.25 \\
FOCES04 & 53099.4230 & $^{\ddag}$ & $^{\ddag}$ & $^{\ddag}$ & $^{\ddag}$ & 5.86 & 6.46 & 7.04 & 6.28 & 6.37 & 6.31 \\
FOCES04 & 53099.4546 & $^{\ddag}$ & $^{\ddag}$ & $^{\ddag}$ & $^{\ddag}$ & 5.92 & 6.34 & 6.95 & 6.36 & 6.38 & 6.27 \\
FOCES04 & 53100.3156 & 6.41 & 6.66 & 6.46 & 5.56 & 6.08 & 6.41 & 6.95 & 6.26 & 6.39 & 6.74 \\
FOCES04 & 53100.3917 & 6.42$^{\dagger}$ & 6.39 & 6.09 & 5.63 & 5.73 & 6.31 & 6.91 & 6.25 & 6.37 & 6.25 \\
FOCES04 & 53101.3264 & $^{\ddag}$ & $^{\ddag}$ & $^{\ddag}$ & 5.84 & 6.14 & 6.33 & 6.91 & 6.22 & 6.36 & 6.26 \\
FOCES04 & 53101.4495 & $^{\ddag}$ & $^{\ddag}$ & $^{\ddag}$ & $^{\ddag}$ & 5.69 & 6.39 & 6.95 & 6.25 & 6.33 & 6.24 \\
FOCES04 & 53102.3306 & 6.55 & 6.47 & 6.23 & $^{\ddag}$ & 6.14 & 6.38 & 7.03 & 6.22 & 6.36 & 6.26 \\
FOCES04 & 53102.4753 & $^{\ddag}$ & $^{\ddag}$ & $^{\ddag}$ & $^{\ddag}$ & $^{\ddag}$ & 6.45 & 6.98 & 6.28 & 6.35 & 6.27 \\
FOCES06 & 54086.6770 & 6.74$^{\blacktriangle}$ & 6.75$^{\blacktriangle}$ &  -   & 5.49 & 5.45  & 5.92 & 6.46 & 5.76 & 6.09 & 6.06 \\
FOCES06 & 54088.5918 & 6.49$^{\blacktriangle}$ & 6.55$^{\blacktriangle}$ & 5.92 & 5.36 & 5.84 & 5.75 & 6.57 & 6.06 & 6.02 & 6.11 \\
FOCES06 & 54091.6192 & 6.68$^{\blacktriangle}$ & 6.65$^{\blacktriangle}$ & 5.87 & 5.50 & 5.53 & 5.81 & 6.49 & 6.07 & 6.29 & 6.05 \\
FOCES07a & 54156.4482 & 6.91$^{\blacktriangle}$ & 6.70$^{\blacktriangle}$ & 6.06 & 5.45 & 5.35 & 5.91 & 6.54 & 6.13 & 6.23 & 6.15 \\
FOCES07a & 54158.5625 & 6.90$^{\blacktriangle}$ & 6.50$^{\blacktriangle}$ & 5.60 &  - & 5.10    & 5.93 & 6.50 & 6.10 & 6.22 & 6.13 \\
FOCES07b & 54228.3461 & -  &  - &  - & - & -   & -  &  6.56 &  6.19 & 6.31 & 6.18  \\
FOCES07b & 54229.3496  & -  &  -  &  -   & - &  - & -  &  6.52 & 6.14 & 6.24 & 6.19 \\
\noalign{\smallskip} \hline
\end{tabular}
\\
{\scriptsize Notes as in previous table}.
\end{center}
\end{flushleft}
\end{table*}

\section[]{Summary and Conclusions}

We have carried out a photometric, polarimetric and spectroscopic
study of FR~Cnc.

Optical $B$-band photometry was carried out during 2007 March --
2008 February at Terskol Branch of the Astronomy Institute
(Russia). There are no peculiar features in the 2008 light curve,
while in the 2007 photometry two brightening episodes were
detected. One of them occurred at the same phase as the flare of
November 23, 2006 (phase = $0.88$) and probably indicates that
both of these events (i.e. the flare on November 23, 2006 and the
photometric brightening episode) originated from the same
long-living active region on FR~Cnc. The non-detection of any
other flares in our photometry except 2006 November 23 implies
that FR~Cnc has a low frequency of flares.

We analysed \emph{ASAS-3} photometry obtained in 2002--2008 in
$V$-band. No evidence of flares in \emph{ASAS-3} data were found.
The profiles of variability are different from season to season.
The mean magnitude in $V$-band remained the same ($V_{\rm mean} =
10.439$ mag) during 2002--2008, while the amplitude decreased
abruptly in 2005. The proposed interpretation is a redistribution
of spots/spot groups over the surface of the star, while the total
percentage of the spotted area was assumed to be constant within
the error limits. A detailed periodogram study of the
\emph{ASAS-3} photometric data enabled us to derive a more
accurate value for the period of FR~Cnc. We find that $P = 0.08265
\pm 0.000015$ d. In addition, we also presented $BVR_{c}I_{c}$
photometric calibration of 166 stars in FR~Cnc vicinity, whose
$V$-magnitude is in the range of $9.85$--$18.06$ mag.

The $BVR$ broad-band polarimetric observations of FR~Cnc have been
obtained at ARIES in Nainital (India) at Manora Peak. The observed
polarization in $B$-band is well matched with the theoretical
values expected for Zeeman polarization model. However, the
observed polarization in $V$ and $R$ bands slightly exceeds the
theoretical values and Thompson and Rayleigh scattering from
inhomogeneous regions are
  not enough to explain the observed polarization excess.
 Therefore the excess of linear polarization should come from an additional source of
 polarization. Taking into account that we conclude that FR~Cnc is not a binary, the mechanism which can produce additional linear polarization is probably scattering in circumstellar material distributed in an asymmetric geometry (e.g. see \citeauthor{Pandey09}, \citeyear{Pandey09}).

A total of 58 spectra of FR~Cnc, which have been obtained  in
2004--2008, were analysed in this work. Based on our spectroscopic
observations, FR~Cnc was classified as K7V star. RV analysis
supports the single nature of FR~Cnc. Anticorrelation between BIS
and RV also indicates that the RV variations are due to stellar
activity variations and not due to a secondary companion.

The kinematics study, based on obtained galactic space-velocity
components ($U,~V,~W$) of FR~Cnc, shows that this star clearly
lies in the young disc population velocity space and might also
belong to IC 2391 moving group, although the Eggen kinematic
criteria shows that FR~Cnc may not be a member of any MG in the
young disc area. The Li~{\sc i} $\lambda$6707.8 averaged $EW$
measured is
 34 m\AA, giving the spectral type of FR~Cnc, it is in agreement
with being a young object between 10--120 Myr.

The H$\alpha$ line was always observed above the continuum in all
the obtained spectra. Measuring the $EW$ of this line, we found
that the H$\alpha$ emission $EW$ average in every season is quite
different. In 2004, as with the photometry, spectroscopic
indicators of chromospheric activity show a high level of activity
which decreased in 2005. The Ca~{\sc ii} (IRT) is included in our
echelle spectra. From the ratio of excess emission $EW$ we found
that in FR~Cnc, Ca~{\sc ii} emission comes from plage-like
regions. We noticed that FR~Cnc can show an activity cycle of 4-5
years, although further follow up
 will confirm this periodicity.

Since FR~Cnc is a rapid rotator, we generated an indirect starspot
map using the Doppler Tomography of Stars imaging code. From it we
derive $v$~sin~$i$ = $46.2 \pm 0.8$ km~s$^{-1}$ and $i = 55 \pm
5^{\circ}$. FR~Cnc belongs to one of the latest spectral types to
have been imaged with the Doppler Tomography.

We independently estimated a rotational velocity of FR~Cnc during
our observations using Queloz method
 (Sect. 5.3) and by the D.I. fits (Sect. 7). Although
 they are differences in the results of the two methods,
 they are consistent. In Table \ref{tab:par} we have put as
 the rotational velocity of FR~Cnc, the value obtained in the D.I.
 as it is likely more accurate.

Despite the short rotation period and its late spectral type, FR~Cnc
shows very few flare events.
 It shows high level of activity as it is a young star, but an unusually
short variability due to the redistribution of activity features
on the stellar surface. While this variability is reflected in the
changes of the amplitude of brightness, the mean brightness
permanently is nearly constant, indicating that the percentage of
stellar surface covered by spots is also constant. The spots
location is also unusual, not showing a polar spot like other F--K
stars do but a distribution more resembling those seen in M1--M2
dwarfs. Although this may be indicative of a distributed dynamo,
the mid-high latitude spot locations are more suggestive of an
interface dynamo under the action of rapid rotation. We can only
speculate as to whether FR~Cnc is representative of a regime in
which a convective-shell-type dynamo gives way to a fully
convective dynamo. Polarimetric observations of the magnetic field
by \citeauthor{donati08mdwarfs} (\citeyear{donati08mdwarfs}) and
\citeauthor{morin08mdwarfs} (\citeyear{morin08mdwarfs}) for
example suggest that this occurs at a later spectral type of M4,
whereas other chromospheric indicators show no obvious changes
until later M spectral types (e.g. \citeauthor{mohanty03activity}
\citeyear{mohanty03activity}). Further spectroscopy with a higher
cadence would enable more detailed maps to be derived, with
multiple epochs enabling the evolution of starspots to be
investigated.

\section*{Acknowledgments}
A. Golovin is thankful to Dr. Anju Mukadam and Dr. Paula Szkody
for useful discussions concerning Fisher Randomization Test in
periodogram analysis, to Dr. Andr\'as Holl (Konkoly Observatory,
Budapest, Hungary) for valuable comments on VOTable format and
help with converting photometric sequence to it in order to align
on DSS image in {\sc aladin}, to Nick Malygin and Dr. Ludmila
Pakuliak for useful discussions and proof-reading of the
manuscript.

. M.C.G\'alvez-Ortiz acknowledges the financial support from the
European Commission
 in the form of Marie Curie Intra European Fellowship (PIEF-GA-2008-220679) and the
 partial support by the Spanish MICINN under the Consolider-Ingenio 2010 Program
 grant CSD2006-00070: First Science with the GTC (http://www.iac.es/consolider-ingenio-gtc).
M.C. G\'alvez-Ortiz and J. Barnes also has received support from
RoPACS during this research, a Marie Curie Initial Training
Network funded by the European Commission's Seventh Framework
Programme. This work was partly supported by the Spanish
Ministerio de Ciencia e Innovacion (MICINN), Programa Nacional de
Astronomia y Astrofisica under grant AYA2008-00695, and under
grant AYA2008-06423-C03-03, and the Comunidad de Madrid under
PRICIT project S2009/ESP-1496 (AstroMadrid). A. Golovin is
thankful to Universidad Complutense de Madrid for hospitality and
for all the efforts and the help during his visit to Spain in 2008
July  and 2009 February.

This publication made use of the {\sc aladin} interactive sky
atlas, operated at CDS, Strasbourg, France (\citeauthor{aladin}
\citeyear{aladin}) and of NASA's Astrophysics Data System.

 \bsp

\label{lastpage}


\begin{thebibliography}{99}
\bibitem[\protect\citeauthoryear{Alekseev}{2003}]{alekseev} Alekseev I.~Y., 2003, ARep, 47, 430

\bibitem[\protect\citeauthoryear{Antoja et~al.}{2008}]{b1b} Antoja T., Figueras F., Fern\'andez D., Torra J., 2008, A\&A, 490, 135
\bibitem[\protect\citeauthoryear{Ar\'evalo~\&~L\'azaro}{1999}]{arevalo99} Ar\'evalo M. J. \& L\'azaro C., 1999, AJ, 118, 1015
\bibitem[\protect\citeauthoryear{Baliunas et~al.}{1996}]{Baliunas} Baliunas S.~L., Nesme-Ribes E., Sokoloff
D., Soon W.~H., 1996, ApJ, 460, 848

\bibitem[\protect\citeauthoryear{Barden}{1985}]{Barden} Barden S.~C., 1985, ApJ, 295, 162

\bibitem[\protect\citeauthoryear{Barnes et~al.}{1998}]{barnes98aper} Barnes J.~R., Collier Cameron A.,
 Unruh~Y.~C., Donati~J.~F., Hussain~G.~A.~J., 1998, MNRAS, 299, 904

\bibitem[\protect\citeauthoryear{Barnes et~al.}{2001}]{barnes01mdwarfs} Barnes J.~R., Collier Cameron A.,
 2001, MNRAS, 326, 950

\bibitem[\protect\citeauthoryear{Barnes et~al.}{2005}]{barnes05lopeg} Barnes J.~R., Cameron A.~C.,
 Lister T.~A., Pointer G.~R., Still M.~D., 2005, MNRAS, 356, 1501

\bibitem[\protect\citeauthoryear{Beavers et~al.}{1979}]{Beavers} Beavers W.~I., Eitter J.~J., Ketelsen D.~A., Oesper D.~A., 1979, PASP, 91, 698

\bibitem[\protect\citeauthoryear{Bessell}{1979}]{Bessell}
Bessell M.~S., 1979, PASP, 91, 589
\bibitem[\protect\citeauthoryear{Bonfils et al.}{2007}]{bonfils07} Bonfils X. et~al., 2007, A\&A, 474, 293
\bibitem[\protect\citeauthoryear{Bonnarel et~al.}{2000}]{aladin} Bonnarel F. et~al., 2000, A\&AS, 143, 33

\bibitem[\protect\citeauthoryear{Breger et~al.}{1993}]{breger93} Breger M. et~al., 1993, A\&A, 271, 482

\bibitem[\protect\citeauthoryear{Breger et~al.}{1999}]{sigma2} Breger M. et~al., 1999, A\&A, 349, 225

\bibitem[\protect\citeauthoryear{Buzasi}{1989}]{Buz89} Buzasi D.~L. 1989, PhD Thesis, Pennsylvania State Univ.

\bibitem[\protect\citeauthoryear{Cameron}{2001}]{cameron01mapping} Collier Cameron A.,
 2001, 'Astrotomography - Indirect Imaging Methods in Observational Astronomy', Springer
 (Lecture Notes in Physics), 183

\bibitem[\protect\citeauthoryear{Chester}{1991}]{chester} Chester M.~M. 1991, PhD Thesis, Pennsylvania State Univ.

\bibitem[\protect\citeauthoryear{Donati et~al.}{1997}]{donati97zdi} Donati J.-F.,
 Semel M., Carter B., Rees D. E., Collier Cameron A., 1997, MNRAS, 291, 658

\bibitem[\protect\citeauthoryear{Donati et~al.}{2008}]{donati08mdwarfs}
Donati J.-F. et al., 2008, MNRAS, 390, 545

\bibitem[\protect\citeauthoryear{Dorren, Guinan
\& Dewarf}{1994}]{dorren} Dorren J.~D., Guinan E.~F., Dewarf
L.~E., 1994, ASPC, 64, 399

\bibitem[\protect\citeauthoryear{Eggen}{1984a}]{Eggen84a} Eggen O.~J., 1984a, AJ, 89, 1358

\bibitem[\protect\citeauthoryear{Eggen}{1984b}]{Eggen84b} Eggen O.~J., 1984b, ApJS, 55, 597

\bibitem[\protect\citeauthoryear{Eggen}{1989}]{Eggen89} Eggen O.~J., 1989, PASP, 101, 366

\bibitem[\protect\citeauthoryear{Elias \& Dorren}{1990}]{elias} Elias N.~M., II, Dorren J.~D., 1990, AJ, 100, 818

\bibitem[\protect\citeauthoryear{Elvis et~al.}{1992}]{intr2} Elvis M., Plummer D., Schachter J., Fabbiano G., 1992, ApJS, 80, 257

\bibitem[\protect\citeauthoryear{ESA}{1997}]{ESA} ESA, 1997, The Hipparcos and Tycho Catalogues, ESA SP-1200

\bibitem[\protect\citeauthoryear{Famaey et~al.}{2007}]{b36b} Famaey B., Pont F., Luri X., Udry S., Mayor M., Jorissen~A., 2007, A\&A, 461, 957
\bibitem[\protect\citeauthoryear{Famaey et~al.}{2008}]{b36c} Famaey B., Siebert A., Jorissen A.,  2008, A\&A, 483, 453

\bibitem[\protect\citeauthoryear{Favata}{1993}]{Favata} Favata F., Barbera M., Micela G., Sciortino S.,
1993, A\&A, 277, 428

\bibitem[\protect\citeauthoryear{Fisher}{1935}]{fisher} Fisher R.A., 1935, Design of Experiments, Oliver and Boyd, Edinburgh.

\bibitem[\protect\citeauthoryear{Fisher}{1936}]{fisher2} Fisher R.A., 1936, J.R.Anthropol.Inst., 66, 57.


\bibitem[\protect\citeauthoryear{G\'alvez et~al.}{2002}]{Galvez02} G\'alvez M.~C. et~al., 2002, A\&A, 389, 524

\bibitem[\protect\citeauthoryear{G\'alvez}{2005}]{Galvez05} G\'alvez M.~C., 2005, PhD Thesis, Universidad Complutense
 de Madrid

\bibitem[\protect\citeauthoryear{G\'alvez et~al.}{2007}]{Galvez07} G\'alvez M.~C. et~al., 2007, A\&A, 472, 587
\bibitem[\protect\citeauthoryear{G\'alvez et~al.}{2009}]{Galvez09} G\'alvez M.~C. et~al., 2009, AJ, 137, 3965
\bibitem[\protect\citeauthoryear{Golovin et~al.}{2007}]{golovin} Golovin A., Pavlenko E., Kuznyetsova Y.,
Krushevska V., 2007, IBVS, 5748, 1

\bibitem[\protect\citeauthoryear{Granzer et~al.}{2000}]{granzer00} Granzer T., Sch{\"u}ssler M.,
 Caligari P., Strassmeier K.~G, 2000, A\&A, 355, 1087

\bibitem[\protect\citeauthoryear{Hall}{1996}]{Hall} Hall
J.~C., 1996, PASP, 108, 313

\bibitem[\protect\citeauthoryear{Hall \& Ramsey}{1992}]{HallRamsey} Hall J.~C., Ramsey L.~W., 1992, AJ, 104, 1942

\bibitem[\protect\citeauthoryear{Huovelin
\& Saar}{1991}]{polar1} Huovelin J., Saar S.~H., 1991, ApJ, 374,
319

\bibitem[\protect\citeauthoryear{Jeffers et~al.}{2007}]{jeffers07abdor} Jeffers S.~V., Donati J.-F., Collier Cameron A., MNRAS, 375, 567

\bibitem[\protect\citeauthoryear{Johnson}{1966}]{johnson} Johnson H.~L., 1966, ARA\&A, 4, 193

\bibitem[\protect\citeauthoryear{Johnson \& Soderblom}{1987}]{JohnsonSoderblom} Johnson D.~R.~H., Soderblom D.~R., 1987, AJ, 93, 864

\bibitem[\protect\citeauthoryear{Kallinger, Reegen
\& Weiss}{2008}]{sigma3} Kallinger T., Reegen P., Weiss W.~W.,
2008, A\&A, 481, 571

\bibitem[\protect\citeauthoryear{Kazarovets et~al.}{1999}]{intr3} Kazarovets~E.~V., Samus~N.~N., Durlevich~O.~V., Frolov~M.~S., Antipin~S.~V., Kireeva~N.~N.,
Pastukhova~E.~N., 1999, IBVS, 4659, 1

\bibitem[\protect\citeauthoryear{Kozhevnikova et~al.}{2006}]{vsp1} Kozhevnikova A.~V., Alekseev I.~Y.,
Heckert P.~A., Kozhevnikov V.~P., 2006, IBVS, 5723, 1

\bibitem[\protect\citeauthoryear{Kupka et~al.}{1999}]{Kupka2} Kupka F., Piskunov N., Ryabchikova T.~A.,
Stempels H.~C., Weiss W.~W., 1999, A\&AS, 138, 119

\bibitem[\protect\citeauthoryear{Kupka et~al.}{2000}]{Kupka1}
Kupka F.~G., Ryabchikova T.~A., Piskunov N.~E., Stempels~H.~C.,
Weiss W.~W., 2000, BaltA, 9, 590

\bibitem[\protect\citeauthoryear{Kurucz}{1993}]{Kurucz}
 Kurucz R.~L. 1993, in Astronomical Society of the Pacific Conference Series,
 Vol.~44, IAU Colloq. 138: Peculiar versus Normal Phenomena in A-type and
 Related Stars, ed. Dworetsky M.~M., Castelli F., Faraggiana R., 87.


\bibitem[\protect\citeauthoryear{Kuschnig et~al.}{1997}]{Kuschnig} Kuschnig R., Weiss W.~W., Gruber R., Bely
P.~Y., Jenkner~H., 1997, A\&A, 328, 544

\bibitem[Lenz
\& Breger(2005)]{period04} Lenz P., Breger M., 2005,
Communications in Asteroseismology, 146, 53

\bibitem[\protect\citeauthoryear{L\'opez-Santiago et~al.}{2003}]{lopez03} L\'opez-Santiago J., Montes D., Fern{\'a}ndez-Figueroa~M.~J., Ramsey L.~W., 2003, A\&A, 411, 489
\bibitem[\protect\citeauthoryear{L\'opez-Santiago et~al.}{2009}]{lopez09} L\'opez-Santiago J., Micela G., Montes D., 2009, A\&A, 499, 129

\bibitem[\protect\citeauthoryear{L\'opez-Santiago et~al.}{2010}]{lopez10} L{\'o}pez-Santiago J., Montes D.,
G{\'a}lvez-Ortiz M.~C., Crespo-Chac{\'o}n I.,
Mart{\'i}nez-Arn{\'a}iz R.~M., Fern{\'a}ndez-Figueroa~M.~J., de
Castro E., Cornide M., 2010, A\&A, 514, A97

\bibitem[\protect\citeauthoryear{Lorente \& Montesinos}{2005}]{lo05} Lorente R., Montesinos B., 2005, AJ, 632, 1104

\bibitem[\protect\citeauthoryear{Maldonado}{2010}]{maldonado10} Maldonado J., Mart\'inez-Arn\'aiz R. M., Eiroa C., Montes D., Montesinos B., 2010, A\&A, 521A, 12

\bibitem[\protect\citeauthoryear{Mamajek
\& Hillenbrand}{2008}]{Mamajek} Mamajek E.~E., Hillenbrand L.~A.,
2008, ApJ, 687, 1264

\bibitem[\protect\citeauthoryear{Mart\'inez-Arn\'aiz et~al.}{2010}]{mart} Mart\'inez-Arn\'aiz R., Maldonado J., Montes D., Eiroa C.,
Montesinos B., 2010, A\&A, 520, A79

\bibitem[\protect\citeauthoryear{Mart\'inez Fiorenzano et~al.}{2005}]{aap} Mart\'inez Fiorenzano A.~F.,
 Gratton R.~G., Desidera S., Cosentino R., Endl M., 2005, A\&A, 442, 775


\bibitem[\protect\citeauthoryear{Medhi et~al.}{2007}]{medhi}
Medhi B.~J., Maheswar G., Brijesh K., Pandey J.~C., Kumar T.~S.,
Sagar R., 2007, MNRAS, 378, 881


\bibitem[\protect\citeauthoryear{Melo et~al.}{2004}]{Melo}
Melo C., Pasquini L., de Medeiros J.~R. 2004, in IAU Symposium,
 Vol. 215, Stellar Rotation, ed. Maeder~A., Eenens~P., 455

\bibitem[{{Mohanty} \& {Basri}(2003)}]{mohanty03activity} Mohanty
S., Basri G., 2003, ApJ, 583, 451

\bibitem[\protect\citeauthoryear{Montes et~al.}{2000}]{Montes00b} Montes D., Fern\'andez-Figueroa M.~J., De Castro E. et~al., 2000, A\&AS, 146, 103

\bibitem[\protect\citeauthoryear{Montes et~al.}{2001}]{Montes01a} Montes D., L\'opez-Santiago J., G\'alvez M.~C. et~al., 2001a,  MNRAS, 328, 45

\bibitem[\protect\citeauthoryear{Montes et~al.}{2001}]{Montes01b}
Montes D., L\'opez-Santiago J., Fern\'andez-Figueroa M.~J.,
G\'alvez M.~C., 2001b, A\&A, 379, 976

\bibitem[\protect\citeauthoryear{Montgomery
\& Odonoghue}{1999}]{sigma1} Montgomery M.~H., Odonoghue D., 1999,
DSSN, 13, 28

\bibitem[\protect\citeauthoryear{Morin et~al.}{2008}]{morin08mdwarfs}
Morin J. et~al., 2008, MNRAS, 390, 567

\bibitem[\protect\citeauthoryear{Noyes et~al.}{1984}]{Noyes}
Noyes R.~W., Hartmann L.~W., Baliunas S.~L., Duncan~D.~K., Vaughan
A.~H., 1984, ApJ, 279, 763

\bibitem[\protect\citeauthoryear{O'Neal, Saar,
\& Neff}{1996}]{1996ApJ...463..766O} O'Neal D., Saar S.~H., Neff
J.~E., 1996, ApJ, 463, 766

\bibitem[O'Neal et al.(1998)]{1998ApJ...507..919O} O'Neal, D., Neff, J.~E.
\& Saar, S.~H.\ 1998, ApJ, 507, 919

\bibitem[O'Neal et al.(2004)]{2004AJ....128.1802O} O'Neal, D., Neff, J.~E.,
Saar, S.~H., \& Cuntz, M.\ 2004, AJ, 128, 1802

\bibitem[\protect\citeauthoryear{Pandey et~al.}{2002}]{intr5} Pandey J.~C., Singh K.~P., Sagar R., Drake S.~A., 2002, IBVS, 5351, 1

\bibitem[\protect\citeauthoryear{Pandey}{2003}]{intr6} Pandey J.~C., 2003, BASI, 31, 329

\bibitem[\protect\citeauthoryear{Pandey et~al.}{2005}]{intr7} Pandey J.~C., Singh K.~P., Drake S.~A., Sagar R., 2005, AJ, 130, 1231

\bibitem[\protect\citeauthoryear{Pandey et~al.}{2009}]{Pandey09} Pandey J.~C., Medhi B.~J., Sagar R., Pandey A.~K., 2009, MNRAS,
396, 1004

\bibitem[\protect\citeauthoryear{Panov, Goranova,
\& Genkov}{2000}]{intr8} Panov K., Goranova Y., Genkov V., 2000,
IBVS, 4917, 1

\bibitem[\protect\citeauthoryear{Pepper et~al.}{2008}]{pepper}
Pepper J., Stanek K.~Z., Pogge R.~W., Latham D.~W., DePoy D.~L.,
Siverd R., Poindexter S., Sivakoff G.~R., 2008, AJ, 135, 907

\bibitem[\protect\citeauthoryear{Perryman et~al.}{1997}]{intr1} Perryman M.~A.~C. et~al., 1997, A\&A, 323, L49

\bibitem[\protect\citeauthoryear{Pojmanski}{2002}]{asas} Pojmanski G., 2002, AcA, 52, 397

\bibitem[\protect\citeauthoryear{Queloz et~al.}{1998}]{Queloz98}
Queloz D., Allain S., Mermilliod~J.-C., Bouvier~J., Mayor~M.
 1998, A\&A, 335, 183

\bibitem[\protect\citeauthoryear{Queloz et~al.}{2001}]{queloz01} Queloz D. et al., 2001, A\&A, 379, 279

\bibitem[\protect\citeauthoryear{Saar
\& Huovelin}{1993}]{polar2} Saar S.~H., Huovelin J., 1993, ApJ,
404, 739


\bibitem[\protect\citeauthoryear{Sbordone et~al.}{2004}]{Sbordone04}
Sbordone L., Bonifacio P., Castelli F., Kurucz R.~L., 2004, MSAIS,
5, 93

\bibitem[\protect\citeauthoryear{Sbordone}{2005}]{Sbordone05}
Sbordone L., 2005, Memorie della Societa Astronomica Italiana
Supplement, 8, 61

\bibitem[\protect\citeauthoryear{Schachter et~al.}{1996}]{intr9} Schachter J.~F., Remillard R., Saar S.~H.,
Favata F., Sciortino S., Barbera M., 1996, ApJ, 463, 747

\bibitem[\protect\citeauthoryear{Schmidt et~al.}{1992}]{schmidt} Schmidt G.~D., Elston R., Lupie O.~L.,
1992, AJ, 104, 1563


\bibitem[\protect\citeauthoryear{Sch\"{u}ssler et~al.}{1996}]{schussler96buoy} Sch\"{u}ssler M.,
 Caligari P., Ferriz-Mas A., Solanki S.~K., Stix~M., 1996, A\&A, 314, 503

\bibitem[\protect\citeauthoryear{Tonry \& Davis}{1979}]{TonryDavis} Tonry J., Davis M., 1979, AJ, 84, 1511

\bibitem[\protect\citeauthoryear{Upgren et~al.}{2002}]{intr10} Upgren A.~R., Sperauskas J., Boyle R.~P.,
2002, BaltA, 11, 91

\bibitem[\protect\citeauthoryear{Voges et~al.}{1999}]{intr11} Voges W. et~al., 1999, A\&A, 349, 389

\bibitem[\protect\citeauthoryear{Zao et~al.}{2009}]{b76b} Zhao J., Zhao G., Chen Y., 2009, ApJ, 692, L113




\end{thebibliography}
\end{document}